\documentclass[12pt, draftclsnofoot, onecolumn]{IEEEtran}
\usepackage[T1]{fontenc}

\usepackage{textcomp}

\usepackage{amsmath,amsfonts,amssymb,amsbsy,amstext}

\usepackage{amsmath}
\usepackage[cmintegrals]{newtxmath}
\usepackage[ruled]{algorithm}
\usepackage{algorithmic}

\usepackage{graphicx}
\usepackage{caption}
\usepackage{makecell}
\usepackage{cite}
\usepackage{multirow}
\usepackage{booktabs}
\usepackage{cite}
\usepackage{tablefootnote}
\usepackage{subfloat}
\usepackage[hyphens]{url}
\usepackage{tikz}
\usetikzlibrary{arrows}
\usetikzlibrary{positioning}
\usetikzlibrary{shapes.misc}
\usetikzlibrary{shapes.symbols}
\usetikzlibrary{shapes.geometric}
\usetikzlibrary{shapes.gates.logic.US}
\usetikzlibrary{decorations.pathreplacing}
\usetikzlibrary{arrows,positioning,matrix,calc,fit}
\usepackage{bm}

\usepackage[display-foreign=false]{acro}

\NewDocumentCommand{\acro}{m o m o}
{%
	\IfValueTF{#2}{%
		\IfValueTF{#4}{%
			\DeclareAcronym{#1}{short={#2},long={#3},#4}
		}{%
			\DeclareAcronym{#1}{short={#2},long={#3}}
		}
	}{%
		\IfValueTF{#4}{%
			\DeclareAcronym{#1}{short={#1},long={#3},#4}
		}{%
			\DeclareAcronym{#1}{short={#1},long={#3}}
		}
	}
}

\acro{1G}{first generation}
\acro{2G}{second generation}
\acro{3G}{third generation}
\acro{4G}{fourth generation}
\acro{5G}{fifth generation}
\acro{5G-StoRM}{5G stochastic radio channel for dual mobility}
\acro{3GPP}{3rd Generation Partnership Project}
\acro{3GPP2}{3rd Generation Partnership Project 2}
\acro{ADMM}{alternating direction method of multipliers}
\acro{BB}{branch and bound}
\acro{BER}{bit error rate}
\acro{BS}{base station}
\acro{CDF}{cumulative distribution function}
\acro{CSI}{channel state information}
\acro{CFN}{channel Frobenius norm}
\acro{DPC}{dirty paper coding}
\acro{DCP}{difference of convex functions program}
\acro{GTEL}{Wireless Telecommunications Research Group}
\acro{GUB}{global upper bound}
\acro{GLB}{global lower bound}
\acro{HMSINR}{highest minimum \ac{SINR}}
\acro{HWCR}{highest \ac{WCR}}
\acro{HWSR}{highest weighted sum rate}
\acro{IBC}{interference broadcast channel}
\acro{IC}{interference channel}
\acro{ITS}{intelligent transport system}
\acro{JSFRA}{joint space-frequency
	resource allocation}
\acro{KKT}{Karush-Kuhn-Tucker}
\acro{LB}{lower bound}
\acro{LOS}{line of sight}
\acro{LP}{linear programming}
\acro{MILP}{mixed integer linear programming}
\acro{MIMO}{multiple input multiple output}
\acro{MINLP}{mixed integer nonlinear programming}
\acro{MIP}{mixed integer programming}
\acro{MISO}{multiple input single output}
\acro{MM}{majorization-minimization}
\acro{MMSE}{minimum mean squared error}
\acro{MRC}{maximum ratio combining}
\acro{MRT}{maximum ratio transmission}
\acro{MSE}{mean squared error}
\acro{MU-MIMO}{multi-user multiple input multiple output}
\acro{MU}{multi-user}
\acro{OFDM}{orthogonal frequency division multiplexing}
\acro{NLP}{nonlinear programming}
\acro{NLOS}{non-line of sight}
\acro{NP}{non-polynomial time}
\acro{PDF}{probability density function}
\acro{RRV}{rate-relaxation variable}
\acro{RUE}{resource usage efficiency}
\acro{QoS}{quality of service}
\acro{SCA}{successive convex approximation}
\acro{SCM}{stochastic channel model}
\acro{SINR}{signal to interference-plus-noise ratio}
\acro{SISO}{single input single output}
\acro{SIMO}{single input multiple output}
\acro{SNR}{Signal to Noise Ratio}
\acro{SOCP}{second-order cone programming}
\acro{TTI}{transmission time interval}
\acro{TU}{typical urban}
\acro{UE}{user equipment}
\acro{UB}{upper bound}
\acro{ULA}{uniform linear array}
\acro{UMi}{urban micro}
\acro{V2X}{vehicle-to-everything}
\acro{WINNER}{Wireless World Initiative New Radio}
\acro{WMMSE}{weighted minimum mean squared error}
\acro{WCR}{weighted common rate}
\acro{WP}{work package}
\acro{ZF}{zero-forcing}
\acro{ZMCSCG}{Zero Mean Circularly Symmetric Complex Gaussian}

\usepackage{siunitx}
\usepackage{datetime}
\usepackage{url}

\usepackage{subfig}

\usepackage[hidelinks=true]{hyperref}
\hypersetup{breaklinks=true}
\newcommand{\FigRef}[2][]{Fig.#1~\ref{#2}}

\usepackage{color}
\definecolor{orange}{rgb}{1,0.5,0}

\newtheorem{prop}{Proposition}

\newtheorem{remark}{Remark}

\DeclareMathAlphabet{\mathppl}{T1}{ppl}{m}{it}
\DeclareMathAlphabet{\mathphv}{T1}{phv}{m}{it}
\DeclareMathAlphabet{\mathpzc}{T1}{pzc}{m}{it}

\DeclareMathOperator{\rank}{rank}

\DeclareMathOperator{\tr}{tr}

\newcommand{\Herm}[1]{{#1}^{\mathrm{H}}}

\newcommand{\Mt}[1]{\mathbf{#1}}

\newcommand{\Rank}[1]{\rank\left(#1\right)}

\newcommand{\Set}[1]{\mathcal{\uppercase{#1}}}

\newcommand{\Trace}[1]{\tr\left(#1\right)}

\newcommand{\Vt}[1]{\mathbf{\lowercase{#1}}}

\newcommand{\mtE}{\Mt{E}}

\newcommand{\mtH}{\Mt{H}}
\newcommand{\mtI}{\Mt{I}}

\newcommand{\mtM}{\Mt{M}}

\newcommand{\mtQ}{\Mt{Q}}

\newcommand{\mtW}{\Mt{W}}
\newcommand{\mtX}{\Mt{X}}

\newcommand{\vtN}{\Vt{N}}

\newcommand{\vtX}{\Vt{X}}
\newcommand{\vtY}{\Vt{Y}}

\newcommand{\stB}{\Set{B}}
\newcommand{\stC}{\Set{C}}

\newcommand{\stN}{\Set{N}}
\newcommand{\stO}{\Set{O}}

\newcommand{\stU}{\Set{U}}

\newcommand{\bbC}{\mathbb{C}}

\newcommand{\bbE}{\mathbb{E}}

\usepackage[final]{fixme}
\fxsetup{theme=color}
\fxuselayouts{marginclue,margin}
\definecolor{fxtarget}{rgb}{0.6000,0.0000,0.6000}
\definecolor{fxerror}{rgb}{0.7300,0.3300,0.8300}

\begin{document}
\bstctlcite{IEEEexample:BSTcontrol}

\listoffixmes

\makeatletter 
\if@twocolumn
\title{Joint Resource Allocation and Transceiver Design for Sum-Rate Maximization under Latency Constraints in Multicell MU-MIMO Systems}
\else 
\title{\huge Joint Resource Allocation and Transceiver Design for Sum-Rate Maximization under Latency Constraints in Multicell MU-MIMO Systems}
\fi \makeatother

\author{Iran M. Braga Jr., Roberto P. Antonioli, G\'abor Fodor, Yuri C. B. Silva,\\ Carlos F. M. e Silva, and Walter C. Freitas Jr.
\thanks{This work was supported in part by Ericsson Research, Technical Cooperation Contract UFC.46 (NAIVE),
in part by the Brazilian National Council for Scientific and Technological Development (CNPq),
in part by the Coordena\c{c}\~{a}o de Aperfei\c{c}oamento de Pessoal de N\'{\i}vel Superior - Brasil (CAPES) - Finance Code 001,
and in part by CAPES/PRINT Grant 88887.311965/2018-00.}
\thanks{Iran M. Braga Jr., Roberto P. Antonioli, Yuri C. B. Silva, Carlos F. M. e Silva
and Walter C. Freitas Jr. are with the  Wireless Telecom Research Group (GTEL),
Federal University of Cear\'a, Fortaleza, Brazil. E-mails: \{iran,antonioli,yuri,cfms,walter\}@gtel.ufc.br.
G\'abor Fodor is with  Ericsson Research and KTH Royal Institute of Technology, Stockholm, Sweden.
(e-mails: gabor.fodor@ericsson.com, gaborf@kth.se).}}
	
\maketitle

\makeatletter 
\if@twocolumn

\begin{abstract}
	Due to the continuous advancements of
	orthogonal frequency division multiplexing (OFDM) and multiple antenna techniques,
	multiuser multiple input multiple output (MU-MIMO) OFDM is a key enabler
	of both fourth and fifth generation networks.
	In this paper, we consider the problem of weighted sum-rate maximization under
	latency constraints in finite buffer multicell MU-MIMO OFDM systems.
	Unlike previous works, the optimization variables include the transceiver beamforming vectors,
	the scheduled packet size and the resources in the frequency and power domains.
	This problem is motivated by the observation that multicell MU-MIMO OFDM systems serve
	multiple quality of service classes and the system performance depends critically
	on both the transceiver design and {the} scheduling algorithm.
	Since this problem is non-convex,
	we resort to the max-plus queuing method and successive convex approximation.
	We propose both centralized and decentralized solutions,
	in which practical design aspects, such as signaling overhead, are considered.
	Finally, we compare the proposed framework with state-of-the-art algorithms
	in relevant scenarios, assuming a realistic channel model with space, frequency and time correlations.
	Numerical results indicate that our design provides significant gains over
	designs based on the wide-spread saturated buffers assumption, while also outperforming algorithms
	that consider a finite-buffer model.
\end{abstract}
\else
\vspace{-7ex}
\begin{abstract}
	Due to the continuous advancements of
	orthogonal frequency division multiplexing (OFDM) and multiple antenna techniques,
	multiuser multiple input multiple output (MU-MIMO) OFDM is a key enabler
	of both fourth and fifth generation networks.
	In this paper, we consider the problem of weighted sum-rate maximization under
	latency constraints in finite buffer multicell MU-MIMO OFDM systems.
	Unlike previous works, the optimization variables include the transceiver beamforming vectors,
	the scheduled packet size and the resources in the frequency and power domains.
	This problem is motivated by the observation that multicell MU-MIMO OFDM systems serve
	multiple quality of service classes and the system performance depends critically
	on both the transceiver design and {the} scheduling algorithm.
	Since this problem is non-convex,
	we resort to the max-plus queuing method and successive convex approximation.
	We propose both centralized and decentralized solutions,
	in which practical design aspects, such as signaling overhead, are considered.
	Finally, we compare the proposed framework with state-of-the-art algorithms
	in relevant scenarios, assuming a realistic channel model with space, frequency and time correlations.
	Numerical results indicate that our design provides significant gains over
	designs based on the wide-spread saturated buffers assumption, while also outperforming algorithms
	that consider a finite-buffer model.
\end{abstract}
\fi \makeatother
	
\begin{IEEEkeywords}
MU-MIMO OFDM, successive convex approximation, QoS, latency, outage probability, multicell.
\end{IEEEkeywords}
	
\IEEEpeerreviewmaketitle
	
\acresetall
	
\section{Introduction}
\label{SEC:INTRODUCTION}
\IEEEPARstart{D}{riven} by the insatiable demands for mobile broadband and low-delay
services, the fourth and fifth generations of cellular networks have been advancing in terms
of capacity, spectral efficiency, reliability and latency.
Indeed, the rapidly increasing number of mobile data subscriptions along with a continuous
increase in the average data volume per subscription has been leading to a
compounded annual growth rate in world-wide data traffic from 8.8 exabytes to above 70
exabytes between 2016 to 2022 \cite{Ericsson}.

In this context, current wireless systems
{do not only have to cope with the growing number of data-hungry applications},
but also with a large {number} of devices connected to the network.
In addition, new use cases, such as intelligent transport {systems},
industrial automation, augmented reality and e-health, impose strict requirements in terms of latency and link reliability, thus demanding sophisticated methods that are able to fulfill these diverse system demands~\cite{Navarro2020,Sutton2019}.
In particular, it is crucial to develop methods that are able to optimize spectrum utilization considering \ac{QoS} constraints.

Due to the continuous advancements of \ac{OFDM} and multiple antenna techniques, \ac{MU-MIMO} \ac{OFDM} has become a key enabler of both fourth and fifth generation (4G/5G) networks. Indeed, \ac{MU-MIMO} systems have the potential to increase the spectral efficiency by allowing multiple independent data streams on shared frequency~\cite{Agiwal2016} while \ac{OFDM} has low-complexity transceiver design, high spectral efficiency, and easy integration with \acs{MIMO} technologies \cite{Zaidi2016}.
	
In general, in \ac{MU-MIMO} \ac{OFDM} systems, binary decision variables
are often advantageously used for frequency-time resource assignment,
which consists in indicating when a given user is assigned to a particular resource
\cite{Yu2012}.  Nevertheless, since linear transmit beamformers are complex vectors, the decision variables can be implicitly modeled by them, which avoids binary decision variables.
In fact, once the design stage is done, the zero transmit beamforming vector indicates
that a user is not assigned on a given resource, whereas the non-zero transmit beamforming vectors
are used to determine the transmission rates of users on a space-frequency resource.
	
In addition, the performance of \ac{MU-MIMO} \ac{OFDM} systems significantly depends
on the considered traffic model.
In this context, two models have been adopted in the recent literature: full-buffer and finite-buffer~\cite{Ameigeiras2012}.
On the one hand, the full-buffer traffic model is characterized by having an unlimited amount of data to transmit in the users' buffers.
Moreover, due to the absence of the packets' arrival process, the number of users in the system is constant.
The main advantage of this model is its simplicity, thus it has been widely adopted for theoretical investigations~\cite{Wei2008,Gao2008}.
On the other hand, the finite-buffer model assumes a finite amount of data
to be transmitted or received in the users' buffers.
Moreover, both packets arrival and departure processes are considered, which leads to a variation in the number of users in the system.
In other works, a given user is assigned a finite payload upon arrival, leaving the system after the payload transmission
or reception is completed.
Even though this model has been less adopted than the full-buffer model, it should be more appropriate for practical scenarios~\cite{Ameigeiras2012}.
	
\section{Related Works and Main Contributions}

Recognizing the importance of resource allocation in multi-antenna systems, the research
community has proposed various resource allocation schemes that {are} applicable specifically in
multi-antenna systems. We categorize these schemes in terms of the number of sub-channels 
and whether they are applicable in full-buffer or finite buffer systems. Following this literature
review, we specify the contribution of the present paper to this line of research.

	\subsection{Resource Allocation in Multi-Antenna Systems with a Single Sub-Channel and Full-Buffer Traffic}
	
	In the literature, transceiver design has {often} been studied
	in the form of optimization problems with different objectives and constraints,
	as seen in~\cite{Castaneda2017} and references therein.
	Among the possible objectives, sum-rate maximization is a well-motivated and often pursued problem formulation.
	In \cite{Oguejiofor2016} the authors studied the sum-rate maximization problem in multicell \ac{MISO} systems, presenting a solution based on the \ac{BB} method. In \cite{Christensen2008} and \cite{Shi2011}, the authors proposed centralized and distributed solutions by exploiting an iterative \ac{WMMSE} approach for a single-cell and multicell \ac{MU-MIMO} systems, respectively.
	Such a scenario was studied in \cite{Shen2018}, for which a centralized iterative algorithm based on fractional programming was proposed.
	In \cite{Codreanu2007} the authors also proposed a centralized solution based on geometric programming.
	An algorithm based on matrix fractional programming was proposed in~\cite{Shen2019}, where the authors proved the convergence of their algorithm using the minorization-maximization approach.
	
	Although these works focused on the weighted sum-rate problem, the scheduling aspect focusing on guaranteeing some per-user minimum \ac{QoS} was not considered. In the literature, \ac{QoS} aspects have been extensively considered in the context of sum-power minimization, such as those in \cite{Shi2016,Tolli2011,Pennanen2016}.
	In~\cite{Shi2016} the sum-power minimization problem subject to \ac{SINR} constraints was studied and a centralized solution based on \ac{MMSE} and uplink-downlink duality was proposed. In addition, a rigorous convergence analysis was also provided.
	Centralized and distributed algorithms for sum-power minimization under minimum \ac{SINR} constrains were also proposed in \cite{Pennanen2016} and \cite{Tolli2011}. Other works, such as \cite{Han2017, Yang2019}, have focused on the sum-energy efficiency maximization problem, that consists in minimizing the energy consumption per transmitted bit.
    The weighted sum-rate maximization problem with \ac{QoS} guarantees was considered in \cite{Kaleva2016,Antonioli2019,Antonioli2020}.
	In~\cite{Kaleva2016} the authors proposed centralized and distributed solutions based on \ac{SCA}, \ac{DCP} and Lagrangian relaxation.
	Centralized, semi-distributed and distributed solutions were proposed in~\cite{Antonioli2019,Antonioli2020} based on the \ac{BB} method, geometric programming and \ac{SOCP}, \ac{SCA} and \ac{DCP}.
	
	\subsection{Resource Allocation in Multi-Antenna Systems with Multiple Sub-Channels and Full-Buffer Traffic}
	Although the works listed in the previous subsection 
	consider some minimum \ac{QoS} assurance, they assume a predetermined frequency-time resource assignment.
	Therefore, the full potential of transceiver design and \ac{MU-MIMO} \ac{OFDM} systems is still not fully explored.
	In the literature, frequency-time resource assignment with \ac{QoS} guarantees in \ac{MU-MIMO} {single-cell} systems was studied in~\cite{Hassan2009,Tejera2009,Lima2014}.
	In \cite{Hassan2009}, the authors proposed a low-complexity frequency-time resource allocation algorithm for the sum-power minimization problem under rate constraints.
	Fairness aspects were considered for \ac{MU-MIMO} \ac{OFDM} in \cite{Tejera2009} by using nonlinear \ac{DPC}-based techniques.
	However, the \ac{DPC} technique makes the proposed solution computationally complex, which is not desirable for practical implementations of real-time networks where complexity is an important feature.
	In~\cite{Lima2012} the sum-rate maximization problem with frequency-time resource assignment was considered in multi-service scenarios. However, the joint space-frequency resource allocation was ignored in~\cite{Lima2012}, which was considered in another work by the same authors in \cite{Lima2014}. Nevertheless, they also assume the use of linear transmit beamformers with equal power allocation, consequently, they do not perform precoder optimization. A joint transceiver design and space-frequency resource allocation {in a multicell} scenario was considered in \cite{Yu2012} and \cite{Yu2013}. In \cite{Yu2012} the authors studied the sum-power minimization problem with \ac{SINR} constraints, where a centralized solution based on a \ac{BB} algorithm was proposed. In \cite{Yu2013}, on the other hand, the weighted sum-rate problem was considered but the \ac{QoS} aspects were ignored. {Moreover, decentralized solutions were not proposed in \cite{Yu2012} and \cite{Yu2013}}.
	
	\subsection{Resource Allocation in Multi-Antenna Systems with Finite-Buffer Traffic}
	{The works listed in the previous subsections have addressed \ac{QoS} aspects and joint space-frequency resource allocation. However, they} assumed a full-buffer model in their modeling. As previously mentioned, these solutions cannot achieve the desired performance in practical scenarios where the finite-buffer model is more suitable. 
	In \cite{Seong2006} the finite-buffer model was considered and the queue minimization problem was studied using geometric programming. In \cite{Mei2018} the authors used the max-plus queuing approach for establishing a suitable model to characterize packet latency and then transform the latency outage probability requirements into minimum data rate constraints. However, these works assumed single antenna transmitters and receivers, which reduces their problems to power allocation problems. In~\cite{Venkatraman2014} the weighted sum-rate problem was studied assuming a finite-buffer model for single-cell \ac{MISO} scenarios. In \cite{Lakshminarayana2013} the authors proposed a distributed solution based on Lyapunov optimization for energy-efficiency with time average \ac{QoS} constraints in a multicell \ac{MISO} scenario.  Centralized and distributed solutions were proposed in \cite{Niu2013} for maximizing the overall system utility while stabilizing all transmission queues in multicell \ac{MU-MIMO} scenarios. {Nevertheless, \cite{Venkatraman2014, Lakshminarayana2013, Niu2013} consider a single sub-channel, thus not exploiting the full potential of \ac{OFDM}.}
	The problem of transceiver design and resource allocation over the space-frequency resources without binary decision variables provided by \ac{MU-MIMO} \ac{OFDM} was studied in~\cite{Venkatraman2016}, where the authors proposed centralized and distributed solutions based on \ac{SCA} and alternating optimization for minimizing the number of backlogged packets waiting at each \ac{BS}.
	In addition, in \cite{Venkatraman2016a} the same authors extended the framework proposed in \cite{Venkatraman2016} in order to reduce the signaling overhead, employing bidirectional training based on over-the-air signaling to update the coupling inter-cell interference variables.
	However, both \cite{Venkatraman2016} and \cite{Venkatraman2016a} ignored per-user \ac{QoS} aspects.
	In \cite{Du2011} the authors considered the latency outage probability requirement, while minimizing \ac{BS} usage and reducing the interfering range in distributed \acs{MIMO} cooperative systems.
	In \cite{Li2013} the weighted sum-rate maximization problem subject to latency outage probability requirements was studied in a single-cell \acs{MU-MIMO} scenario. {However, by increasing the number of cells, a more complex structure of the transceiver must be designed to combat the multicell interference, and the centralized scheme developed in \cite{Li2013} becomes impractical in such scenarios.} 
	
	\subsection{Main Contributions}
	
	
		
	The main contributions of this work are summarized as:
	\begin{enumerate}
		\item Investigation of a variant of the weighted sum-rate maximization problem in which we introduce latency outage probability constraints for multicell \ac{MU-MIMO} \ac{OFDM} systems with finite-buffer. Therefore, the formulated problem considers altogether the transceiver design in \ac{MIMO} systems, user scheduling across multiple sub-channels, minimum \ac{QoS} requirements and a finite-buffer traffic model.
		\item Development of a centralized solution based on \ac{SCA}, which is employed to handle the non-convexity of the formulated optimization problem.
		Moreover, a novel decentralized solution based on a partial Lagrangian relaxation and a subsequent primal-dual method is also provided, for which we present a signaling scheme for deployments in practical multicell \ac{MU-MIMO} \ac{OFDM} systems. The decentralized algorithm is highly non-trivial due to the inherent coupling between the BSs.
		\item A convergence analysis is provided proving that both centralized and decentralized solutions converge to a \ac{KKT} point of the novel formulated problem. Our converge analysis is based on~\cite{Kaleva2016}.
		\item Performance evaluation by means of simulations, where we compare the proposed solution with state-of-the-art optimization-based algorithms {considering different aspects, including Poisson and bursty traffic models, as well as imperfect \ac{CSI}}.
		Unlike previous works, we consider a realistic channel modeling based on the \ac{3GPP}'s stochastic channel model with spatial, frequency and time correlations. 
		Simulations show that the proposed solution efficiently handles the resource allocation across multiple sub-channels and finite-buffer scheduling, while presen\-ting performance gains in terms of outage probability and latency.
	\end{enumerate}
	
	\textit{Organization:} The remainder of the paper is organized as follows.
	Sections~\ref{SEC:SYS_MODEL} and \ref{SEC:PROB_FORM} introduce the system model
	and the weighted sum-rate maximization problem with maximum outage probability constraints, respectively. Section \ref{SEC:CENTRALIZED_SOLUTION} describes the proposed centralized solution.
	Section \ref{SEC:DECENTRALIZED_SOLUTION} presents the proposed decentralized solution along with the involved signaling aspects.
	Section \ref{SEC:PERFORMANCE_EVALUATION} provides the numerical results along with discussions and, finally, Section \ref{SEC:CONCLUSIONS} highlights the main conclusions, as well as perspectives for future works.
	
	\textit{Notation:} Throughout the paper, matrices and vectors are presented by boldface upper and lower case letters, respectively.
	$\mathbf{X}^T$, $\mathbf{X}^H$, $\mathbf{X}^{-1}$ and $\mathbf{X}^\dagger$ stand, respectively, for transpose, Hermitian, inverse and pseudo-inverse of a matrix $\mathbf{X}$.
	$\{x_i\}_{\forall i}$ denotes the set of elements $x_i$ for the values of $i$ denoted by the subscript expression.
	$\mathbf{I}$ is the identity matrix.
	Mapping of negative scalars to zero is written as $(\cdot)^+ = \text{max}(0,\cdot)$.
	Expected value of a random variable is denoted by $\mathbb{E}[\cdot]$.
	{For a matrix $\mtX$, $\Trace{\mtX}$ and $\det(\mtX)$ are the trace and determinant operators, respectively.} 
	
	\section{System Model}
	\label{SEC:SYS_MODEL}
	
	We consider the downlink of a multicell {\acs{MU-MIMO}} scenario in an \ac{OFDM} framework composed of $N$ sub-channels, where $B$ \acfp{BS} equipped with $N_\text{T}$ antennas serve a total of $U$ multi-antenna \acp{UE}, each one equipped with $N_\text{R}$ antennas.
	Let $\stU = \{1,2,\ldots,U\}$ indicate the set of all users in the system.
	The number of users associated with \ac{BS} $b$ is denoted by $U_b$, where each user $u$ is served by a single \ac{BS} $b_u$.
	We assume that all \acp{BS} serve the respective users with linear transmit beamforming.
	
	{We let $S$ denote the maximum number of spatial streams\footnote{{Note that $S \leq \Rank{H_{b_{u},u,n}} = \min(N_{\text{R}}, N_{\text{T}})$. Moreover, the number of streams allocated to each user will be computed by the proposed algorithms, where a zero power transmit beamformer is used for a specific non-activated stream.}}. The downlink signal received by user $u$ on sub-channel $n$ is given by 
	\makeatletter 
	\if@twocolumn 
	\begin{align}
	\vtY_{u,n} = \; \mtH_{b_u,u,n} \mtM_{u,n} \vtX_{u,n} + \sum_{\substack{i=1\\i\neq u}}^{U}\mtH_{b_i,u,n}\mtM_{i,n} \vtX_{i,n} + \vtN_{u,n},
	\end{align}
	\else 
	\begin{align}
		\vtY_{u,n} = \; \mtH_{b_u,u,n} \mtM_{u,n} \vtX_{u,n} + \sum_{\substack{i=1\\i\neq u}}^{U}\mtH_{b_i,u,n}\mtM_{i,n} \vtX_{i,n} + \vtN_{u,n},
	\end{align}
	\fi \makeatother
	where $\textbf{H}_{b_i,u,n} \in \mathbb{C}^{N_\text{R} \times N_\text{T}}$ is the channel matrix between user $u$ and \ac{BS} $b$ serving user $i$ on sub-channel $n$,
	$\mtM_{u,n} \in \mathbb{C}^{N_\text{T}\times S}$ is the transmit beamforming matrix that the \ac{BS} uses on sub-channel $n$ to transmit the symbol $\vtX_{u}\in\bbC^{S\times 1}$ to user $u$ with ${\mathbb{E} \left[\vtX_{u}\Herm{\vtX}_{u}\right] = \mtI}$ and $\textbf{n}_{u,n} \in \mathbb{C}^{N_\text{R}} \sim \stC\stN(0,\sigma^2)$ is the noise at user ${u}$ and sub-channel $n$.
	User $u$ decodes the signal $\textbf{y}_{u,n}$ via a receive beamformer matrix~$\mtW_{u,n}\in\mathbb{C}^{N_\text{R}\times S}$ so that the estimated symbol is given by
	\begin{equation}
		\hat{\vtX}_{u,n} = \Herm{\mtW}_{u,n}\vtY_{u,n}.
	\end{equation}}
	
{Furthermore, the rate assigned to user $u$ is $r_{u} = \sum\limits_{n=1}^{N}r_{u,n}$, where $r_{u,n}$ is the number of transmitted bits per second for user $u$ on sub-channel $n$, which is given by
\begin{equation}
	\label{eq:rate2}
	r_{u,n} = \log_2 \det\left(\mtI + \mtH_{b_{u},u,n}\mtM_{u,n}\Herm{\mtM}_{u,n}\Herm{\mtH}_{b_{u},u,n}\left(\sum_{\substack{i=1\\i\neq u}}^{U}\mtH_{b_{i},u,n}\mtM_{i,n}\Herm{\mtM}_{i,n}\Herm{\mtH}_{b_{i},u,n} + \sigma^2\mtI\right)^{-1}\right).
\end{equation}}
 	
We assume the availability of perfect \ac{CSI} at the transmitters and receivers for the design of the proposed algorithms, similarly to the assumptions used in~\cite{Shi2011,Christensen2008,Kaleva2016,Pennanen2016}. In other words, we assume that the channel matrix is perfectly known at transmitters and receivers without channel estimation errors. Moreover, we consider that the channel matrices are generated using a realistic channel modeling based on the \ac{3GPP}'s stochastic channel model with spatial, frequency and time correlations \cite{Pessoa2019, 3GPP2018Study}.
		
	It is assumed that the packet arrival process for the $u$-th user is independent and identically distributed (i.i.d) over the time slots and follows a Poisson distribution\footnote{We remark that the Poisson model 	is still relevant in practice as it is used for evaluation purposes in 3GPP analyses, such as in \cite{3GPP2017Further,3GPPTS37985,3GPPTR36885}. Furthermore, the Poisson traffic model	is useful because it allows {handling} non-convex optimization problems	and {developing} closed-form equations for centralized and decentralized algorithms.} with average arrival rate $\lambda_u$~\cite{Venkatraman2016,Mei2018}.
	Also, the $m$-th packet size for user $u$, denoted by $L_{u,m}$, is i.i.d over the time slots and follows an exponential distribution with mean packet size $\overline{L}_{u,m}$.
	We let $Q_u$ represent the number of backlogged packets destined for user $u$.
	The waiting time of the $m$-th packet in the buffer of user $u$ is given by $W_{u,m}$ and the respective transmission time is $\delta_{u,m}$.
	Therefore, the latency of the $m$-th packet destined to user $u$ is written as
	\begin{equation}
	\label{eq:latency}
	D_{u,m} = W_{u,m} + \delta_{u,m},
	\end{equation}
	which is given in time slots.
	Let us denote the maximum tolerable latency for packet transmission as $d_\text{max}$ (in time slots) and the maximum outage probability as $\xi$.
	Thus, the latency outage probably requirement of user $u$ can be expressed as
	\begin{equation}
	\label{eq:latency_req}
	P\{D_{u,m}>d_\text{max}\} \leq \xi.
	\end{equation}

	\section{Problem Formulation}
	\label{SEC:PROB_FORM}
	We consider the following variant of the transceiver design problem for weighted sum-rate maximization in \ac{MU-MIMO} \ac{OFDM} systems under per-\ac{BS} maximum power and per-user latency constraints:
	{\makeatletter 
	\if@twocolumn 
\begin{subequations}
	\label{PROB:MAX_RATE_LATENCY}
	\begin{align}\hspace{-0.4cm}
		\underset{\mtW_{u,n},\mtM_{u,n}}{\text{maximize}}  & \quad \sum\limits_{u=1}^U \beta_{u}\sum_{n=1}^{N}r_{u,n} \label{PROB:MAX_RATE_LATENCY_OBJ} \\
		\mbox{subject to} & \quad
		\sum\limits_{u=1}^{U_b}\sum\limits_{n=1}^N \Trace{\mtM_{u,n}\Herm{\mtM}_{u,n}} \leq P_{b}, & \forall b, \label{PROB:MAX_RATE_LATENCY_CONS1} \\
		& \quad P\{D_{u,m}>d_\text{max}\} \leq \xi, & \forall u,m, \label{PROB:MAX_RATE_LATENCY_CONS3}\\
		& \quad \sum_{n=1}^{N}r_{u,n} \leq \frac{Q_u}{\Delta_\text{TTI}}, & \forall u,\label{PROB:MAX_RATE_LATENCY_CONS4}
	\end{align}
\end{subequations}
	\else 
\begin{subequations}
	\label{PROB:MAX_RATE_LATENCY}
	\begin{align}\hspace{-0.4cm}
		\underset{\mtW_{u,n},\mtM_{u,n}}{\text{maximize}}  & \quad \sum\limits_{u=1}^U \beta_{u}\sum_{n=1}^{N}r_{u,n} \label{PROB:MAX_RATE_LATENCY_OBJ} \\
		\mbox{subject to} & \quad
		\sum\limits_{u=1}^{U_b}\sum\limits_{n=1}^N \Trace{\mtM_{u,n}\Herm{\mtM}_{u,n}} \leq P_{b}, & \forall b, \label{PROB:MAX_RATE_LATENCY_CONS1} \\
		& \quad P\{D_{u,m}>d_\text{max}\} \leq \xi, & \forall u,m, \label{PROB:MAX_RATE_LATENCY_CONS3}\\
		& \quad \sum_{n=1}^{N}r_{u,n} \leq \frac{Q_u}{\Delta_\text{TTI}}, & \forall u,\label{PROB:MAX_RATE_LATENCY_CONS4}
	\end{align}
\end{subequations}
	\fi \makeatother}

\noindent where $\beta_{u}>0$ denotes the priority weight of user $u$, $P_b$ denotes the power budget of \ac{BS} $b$ and $\Delta_\text{TTI}$ is the duration of one \ac{TTI}. The optimization variables are the transmit and receive beamforming matrices {$\mtM_{u,n} \in \mathbb{C}^{N_\text{T}\times S}$ and $\mtW_{u,n} \in \mathbb{C}^{N_\text{R}\times S}$ $\forall (u,n)$, respectively}. Observe that constraints~\eqref{PROB:MAX_RATE_LATENCY_CONS3} implicitly depend on the optimization variables. In fact, the higher is the rate allocated to user $u$, the faster the $m$-th packet destined to user $u$ will be transmitted from the \ac{BS} to user $u$. Since the rate allocated to user $u$ depends on the transmit beamformers {$\{\mtM_{u,n}\}_{\forall (n)}$} and on the receive beamformers {$\{\mtW_{u,n}\}_{\forall (n)}$} computed for that user, consequently, both transmit and receive beamforming have an impact on the outage probability requirement of user~$u$. Finally, constraints~\eqref{PROB:MAX_RATE_LATENCY_CONS4} state that the sum of the bits transmitted to user~$u$ cannot be higher than the amount of bits in its buffer in order to avoid excessive allocation of the resources. Unlike the full-buffer model, in which the transmit buffers relative to the users always have an unlimited amount of data to be transmitted, the finite-buffer model assumes that such buffers have a limited amount of data. Thus, the maximum allowed user rate can also be determined by the users' transmit buffer located at the \ac{BS}. Recognizing this, and in order to arrive at industrially applicable results, we are motivated to explicitly incorporate the finite buffer in our model, whose size is a system parameter. The importance of these additional constraints will be clearly shown by means of simulations in Section~\ref{SEC:PERFORMANCE_EVALUATION}.
	
	Therefore, the formulated problem allows to simultaneously optimize the transceiver design and schedule the users across space-frequency resources in order to satisfy their packet latency and rate demands, while also handling more realistic finite-buffer traffic models. However, the packet latency as defined in~\eqref{eq:latency} is very difficult to compute directly, thus the latency constraints \eqref{PROB:MAX_RATE_LATENCY_CONS3} require some transformation so that we arrive at a tractable form. 	
	Besides that, we remark that the sum-rate maximization problem considering an interference channel (i.e., with only one user per \ac{BS}) was shown to be strongly \ac{NP}-hard in Theorem 1 and Theorem 6 of~\cite{Luo2008}. In fact, this \ac{NP}-hardness was demonstrated in~\cite{Luo2008} even for a very simple case (considering only one sub-channel and optimizing only the powers at the \acp{BS}). The sum-rate maximization considered in our case, i.e., problem~\eqref{PROB:MAX_RATE_LATENCY}, has clearly a wider set of feasible solutions, since it involves transceiver designs, power allocation for multiple users per cell, and multiple sub-channels.
	Nevertheless, it is important to realize that adding further constraints to an \ac{NP}-hard problem does not necessarily result in another \ac{NP}-hard problem. Therefore, having a complete proof about the possible NP-hardness of problem \eqref{PROB:MAX_RATE_LATENCY} requires a much more detailed analysis, which is beyond the scope of our work.
	
	However, it is worth highlighting that due to the nonconvexity of problem \eqref{PROB:MAX_RATE_LATENCY}, obtaining its global optimal solution is computationally very difficult. As far as we know, no technique is able to find the optimal solution for problem \eqref{PROB:MAX_RATE_LATENCY}. Motivated by this issue, we develop centralized and decentralized solutions that are capable of computing local optimal solutions to problem \eqref{PROB:MAX_RATE_LATENCY}.
		
	\section{Centralized Solution}
	\label{SEC:CENTRALIZED_SOLUTION}
	This section proposes a centralized solution for problem~\eqref{PROB:MAX_RATE_LATENCY}.
	To this end, we first reformulate the latency constraints in a tractable form. The resulting problem is non-convex, which is then reformulated and iteratively solved by means of \ac{SCA}.
	
	\subsection{Latency Constraint Reformulation}
	
	The main challenge involved in computing the latency outage probability defined in~\eqref{eq:latency_req} lies in the difficulty to calculate~\eqref{eq:latency} using a closed form equation.
	To handle this issue, we resort to the max-plus queuing approach from random network calculation~\cite{Jiang2010,Mei2018}, which enables the transformation of the latency constraint in~\eqref{eq:latency_req} into a data rate constraint.
	
	\begin{prop}
		For each user $u$, when its buffer is not empty at time instant $t$ (i.e., $Q_u(t)>0$), its instantaneous rate $r_u(t)$ must be larger than or equal to the minimum data rate $R_u^\text{min}$ to ensure that the maximum tolerable latency constraint in~\eqref{eq:latency_req} is met. Mathematically, at each time instant~$t$,
		\begin{align}
		\label{eq:r_min_req}
		r_u(t) \begin{cases} \geq R_u^\text{min}, \quad & Q_u(t)>0, \\ = 0, & Q_u(t)=0, \end{cases}
		\end{align}
		where
		\makeatletter 
		\if@twocolumn
		\begin{align}
		\label{eq:r_min}
		R_u^\text{min} = - \frac{\overline{L}_{u,m}}{d_\text{max}}\left[W_{-1}\left(\frac{\lambda_ud_\text{max}\xi}{1-e^{\lambda_ud_\text{max}}}e^{\left(\frac{\lambda_ud_\text{max}}{1-e^{\lambda_ud_\text{max}}}\right)}\right) \right. \nonumber \\ + \left.\left(\frac{\lambda_ud_\text{max}}{1-e^{\lambda_ud_\text{max}}}\right) \right],
		\end{align}
		\else
		
		\begin{align}
		\label{eq:r_min}
		R_u^\text{min} = - \frac{\overline{L}_{u,m}}{d_\text{max}}\left[W_{-1}\left(\frac{\lambda_ud_\text{max}\xi}{1-e^{\lambda_ud_\text{max}}}e^{\left(\frac{\lambda_ud_\text{max}}{1-e^{\lambda_ud_\text{max}}}\right)}\right)  + \left(\frac{\lambda_ud_\text{max}}{1-e^{\lambda_ud_\text{max}}}\right) \right],
		\end{align}
		\fi \makeatother
		where $W_{-1}(x) : x \in [-e^{-1},0] \rightarrow [-\infty,-1]$ is the lower branch of the Lambert function $W$ satisfying $z=W_{-1}(ze^z)$.
	\end{prop}
	\begin{IEEEproof}
		The proof of this theorem follows from \textit{Lemma 1} and \textit{Theorem 1} from~\cite{Mei2018}.
	\end{IEEEproof}	
	
	Using Proposition~1, we can replace the latency requirement constraints~\eqref{PROB:MAX_RATE_LATENCY_CONS3} by the minimum rate constraint~\eqref{eq:r_min_req}.
	Furthermore, considering the set $\stU_t = \{u\;|\;Q_u(t)>0, \, u \in \stU\}$, which is the set of users with bits to be received from the \ac{BS}, problem~\eqref{PROB:MAX_RATE_LATENCY} can be reformulated as follows:
	{\makeatletter 
	\if@twocolumn 
	\begin{subequations}
		\label{PROB:MAX_RATE_LATENCY_REF}
		\begin{align}\hspace{-0.4cm}
			\underset{\mtW_{u,n}, \mtM_{u,n}}{\text{maximize}} & \quad \sum\limits_{u=1}^U \beta_{u}\sum_{n=1}^{N}r_{u,n} \label{PROB:MAX_RATE_LATENCY_REF_OBJ}\\
			\mbox{subject to}
			& \quad \sum_{n=1}^{N}r_{u,n} \geq R_u^\text{min} , & u \in \stU_t, \label{PROB:MAX_RATE_LATENCY_REF2}\\
			& \quad \eqref{PROB:MAX_RATE_LATENCY_CONS1}, \text{ and } \eqref{PROB:MAX_RATE_LATENCY_CONS4},
		\end{align}
	\end{subequations}
	\else 
	\begin{subequations}
		\label{PROB:MAX_RATE_LATENCY_REF}
		\begin{align}\hspace{-0.4cm}
			\underset{\mtW_{u,n}, \mtM_{u,n}}{\text{maximize}} & \quad \sum\limits_{u=1}^U \beta_{u}\sum_{n=1}^{N}r_{u,n} \label{PROB:MAX_RATE_LATENCY_REF_OBJ}\\
			\mbox{subject to}
			& \quad \sum_{n=1}^{N}r_{u,n} \geq R_u^\text{min} , & u \in \stU_t, \label{PROB:MAX_RATE_LATENCY_REF2}\\
			& \quad \eqref{PROB:MAX_RATE_LATENCY_CONS1}, \text{ and } \eqref{PROB:MAX_RATE_LATENCY_CONS4},
		\end{align}
	\end{subequations}
	\fi \makeatother}

\noindent where this problem is solved for each time instant $t$.
	Problem~\eqref{PROB:MAX_RATE_LATENCY_REF} is nonconvex, which makes the global optimal solution computationally difficult to be obtained. Therefore, in order to design computationally lower complexity and practical solutions while preserving an efficient performance, we resort to an approximation approach.
	
	Due to rate and power constraints, problem \eqref{PROB:MAX_RATE_LATENCY_REF} can be infeasible.  In fact, the fulfillment of rate constraints in interference-limited systems can cause feasibility issues even without power constraints. Meanwhile, the restricted power budget can render the problem infeasible when the rate constraints are set too high.
		
	\subsection{Problem Reformulation}
	
	Considering the users' viewpoint, the well-known linear \ac{MMSE} receiver is the rate optimal linear receiver since it maximizes the per-stream \ac{SINR} and, consequently, the per-user rate~\cite{Kaleva2016, Pennanen2016}. {The matrix expression for the \ac{MMSE} receiver of user $u$ on sub-channel $n$ is 
	\makeatletter 
	\if@twocolumn
	\begin{equation}
		\begin{split}
			\mtW_{u,n} = \Bigg(\sum_{i=1}^{U}\mtH_{b_{i},u,n}\mtM_{i,n}\Herm{\mtM}_{i,n}\Herm{\mtH}_{b_{i},u,n} + \sigma^2\mtI\Bigg)^{-1}
			\mtH_{b_{u},u,n}\mtM_{u,n}, \label{EQ:MMSE}
		\end{split}
	\end{equation}
	\else 
	\begin{equation}
		\begin{split}
			\mtW_{u,n} = \Bigg(\sum_{i=1}^{U}\mtH_{b_{i},u,n}\mtM_{i,n}\Herm{\mtM}_{i,n}\Herm{\mtH}_{b_{i},u,n} + \sigma^2\mtI\Bigg)^{-1}
			\mtH_{b_{u},u,n}\mtM_{u,n}, \label{EQ:MMSE}
		\end{split}
	\end{equation}
	\fi \makeatother
	and the \ac{MSE} matrix for user $u$ on sub-channel $n$ is given by
	\makeatletter 
	\if@twocolumn
	\begin{align}
		\mtE_{u,n} &= \bbE\left[(\Herm{\mtW}_{u,n}\vtY_{u,n} - \vtX_{u,n})\Herm{(\Herm{\mtW}_{u,n}\vtY_{u,n} - \vtX_{u,n})}\right]\nonumber\\
		&= \left(\mtI - \Herm{\mtW}_{u,n}\mtH_{b_{u},u,n}\mtM_{u,n}\right)\Herm{\left(\mtI - \Herm{\mtW}_{u,n}\mtH_{b_{u},u,n}\mtM_{u,n}\right)}\nonumber \\
		&\hspace{10ex} + \sum_{\substack{i=1\\i\neq u}}\Herm{\mtW}_{u,n}\mtH_{b_{i},u,n}\mtM_{i,n}\Herm{\mtM}_{i,n}\Herm{\mtH}_{b_{i},u,n}\mtW_{u,n} + \sigma^2\Herm{\mtW}_{u,n}\mtW_{u,n}. \label{EQ:MSE}
	\end{align}
	\else
	\begin{align}
		\mtE_{u,n} &= \bbE\left[(\Herm{\mtW}_{u,n}\vtY_{u,n} - \vtX_{u,n})\Herm{(\Herm{\mtW}_{u,n}\vtY_{u,n} - \vtX_{u,n})}\right]\nonumber\\
		&= \left(\mtI - \Herm{\mtW}_{u,n}\mtH_{b_{u},u,n}\mtM_{u,n}\right)\Herm{\left(\mtI - \Herm{\mtW}_{u,n}\mtH_{b_{u},u,n}\mtM_{u,n}\right)}\nonumber \\
		&\hspace{10ex} + \sum_{\substack{i=1\\i\neq u}}\Herm{\mtW}_{u,n}\mtH_{b_{i},u,n}\mtM_{i,n}\Herm{\mtM}_{i,n}\Herm{\mtH}_{b_{i},u,n}\mtW_{u,n} + \sigma^2\Herm{\mtW}_{u,n}\mtW_{u,n}. \label{EQ:MSE}
	\end{align}
	\fi \makeatother}

	{Now, by assuming that \ac{MMSE} receivers are employed at all users, 
	we take advantage of a useful relation between the \ac{MSE}, $\mtE_{u,n}\in\bbC^{S\times S}$, and the rate,  $r_{u,n}$~\cite{Christensen2008,Shi2011}:
	\begin{equation}
		r_{u,n} = \log_2\det\left(\mtE_{u,n}^{-1}\right). \label{EQ:MSE_SINR}
	\end{equation}}
	
	{At this point, we can replace \eqref{EQ:MSE_SINR} in \eqref{PROB:MAX_RATE_LATENCY_REF}, use the relaxed \ac{MSE} expression in \eqref{EQ:MSE} and apply the relaxed rate expression $\hat{r}_{u,n} \leq r_{u,n}$, which facilitates the \ac{SCA} that will be later adopted. We can thus reformulate the original problem as follows:
	\makeatletter 
	\if@twocolumn
	\begin{subequations}
	\label{EQ:PROB_MSE}
	\begin{align}\hspace{-0.4cm}
		\underset{\substack{\mtW_{u,n},\mtM_{u,n},\\
				\hat{r}_{u,n},\mtE_{u,n}}}{\text{maximize}} & \quad \sum\limits_{u=1}^U \beta_{u}\sum\limits_{n=1}^N  \hat{r}_{u,n} \label{EQ:PROB_MSE_OBJ}\hspace{-1.5ex}\\
		\mbox{subject to}
		& \quad \hat{r}_{u,n} \leq -\log_2\det(\mtE_{u,n}),&  \forall u\in\stU_t,n,\label{EQ:PROB_MSE_CONS2}\\
		& \quad \sum\limits_{n=1}^N \hat{r}_{u,n} \geq R_u^\text{min},&  u \in \stU_t,\label{EQ:PROB_MSE_CONS3}\\
		& \quad \sum\limits_{n=1}^N \hat{r}_{u,n} \leq \frac{Q_u}{\Delta_\text{TTI}},\ &  u \in \stU_t, \label{EQ:PROB_MSE_CONS4}\\
		& \quad \left(\mtI - \Herm{\mtW}_{u,n}\mtH_{b_{u},u,n}\mtM_{u,n}\right)\Herm{\left(\mtI - \Herm{\mtW}_{u,n}\mtH_{b_{u},u,n}\mtM_{u,n}\right)}\nonumber \\
		& + \sum_{\substack{i=1\\i\neq u}}\Herm{\mtW}_{u,n}\mtH_{b_{i},u,n}\mtM_{i,n}\Herm{\mtM}_{i,n}\Herm{\mtH}_{b_{i},u,n}\mtW_{u,n} + \sigma^2\Herm{\mtW}_{u,n}\mtW_{u,n} \leq \mtE_{u,n},& \forall u\in\stU_t, n,\label{EQ:PROB_MSE_CONS5} \\
		& \quad \text{ and } \eqref{PROB:MAX_RATE_LATENCY_CONS1}.
	\end{align}
\end{subequations}
	\else
	\begin{subequations}
	\label{EQ:PROB_MSE}
	\begin{align}\hspace{-0.4cm}
		\underset{\substack{\mtW_{u,n},\mtM_{u,n},\\
				\hat{r}_{u,n},\mtE_{u,n}}}{\text{maximize}} & \quad \sum\limits_{u=1}^U \beta_{u}\sum\limits_{n=1}^N  \hat{r}_{u,n} \label{EQ:PROB_MSE_OBJ}\hspace{-1.5ex}\\
		\mbox{subject to}
		& \quad \hat{r}_{u,n} \leq -\log_2\det(\mtE_{u,n}),&  \forall u\in\stU_t,n,\label{EQ:PROB_MSE_CONS2}\\
		& \quad \sum\limits_{n=1}^N \hat{r}_{u,n} \geq R_u^\text{min},&  u \in \stU_t,\label{EQ:PROB_MSE_CONS3}\\
		& \quad \sum\limits_{n=1}^N \hat{r}_{u,n} \leq \frac{Q_u}{\Delta_\text{TTI}},\ &  u \in \stU_t, \label{EQ:PROB_MSE_CONS4}\\
		& \quad \left(\mtI - \Herm{\mtW}_{u,n}\mtH_{b_{u},u,n}\mtM_{u,n}\right)\Herm{\left(\mtI - \Herm{\mtW}_{u,n}\mtH_{b_{u},u,n}\mtM_{u,n}\right)}\nonumber \\
		& + \sum_{\substack{i=1\\i\neq u}}\Herm{\mtW}_{u,n}\mtH_{b_{i},u,n}\mtM_{i,n}\Herm{\mtM}_{i,n}\Herm{\mtH}_{b_{i},u,n}\mtW_{u,n} + \sigma^2\Herm{\mtW}_{u,n}\mtW_{u,n} \leq \mtE_{u,n},& \forall u\in\stU_t, n,\label{EQ:PROB_MSE_CONS5} \\
		& \quad \text{ and } \eqref{PROB:MAX_RATE_LATENCY_CONS1}.
	\end{align}
\end{subequations}
	\fi \makeatother}

	Problem \eqref{EQ:PROB_MSE} is still a non-convex problem even for fixed receive beamformers, {$\{\mtW_{u,n}\}_{\forall (u\in\stU_t,n)}$}. 
	Fortunately, following the approach from~\cite{Venkatraman2016}, we can resort to the \ac{SCA} approach \cite{Boyd2007,Marks1978} to relax the non-convex rate constraints in~\eqref{EQ:PROB_MSE_CONS2} using a sequence of convex subsets by applying the first-order Taylor approximation around a fixed \ac{MSE} point {$\mtE^{(k)}_{u,n}$ as
	\begin{equation}
		\hat{r}_{u,n} \leq -\log_2\det\left(\mtE^{(k)}_{u,n}\right) -  \frac{\Trace{\left(\mtE_{u,n}^{(k)}\right)^{-1}\left(\mtE_{u,n} - \mtE_{u,n}^{(k)}\right)}}{\log(2)}, \label{EQ:TAYLOR}
	\end{equation}
	where $\{\mtE^{(k)}_{u,n}\}_{\forall(u\in\stU_t,n)}$} denotes the points of approximation for the spatial data streams in the $k$th iteration. Using \eqref{EQ:TAYLOR}, the rate constraints in~\eqref{EQ:PROB_MSE_CONS2} can be approximated as convex constraints. Finally, problem \eqref{EQ:PROB_MSE} can be presented in the $k$th iteration for fixed {$\{\mtE^{(k)}_{u,n}\}_{\forall(u\in\stU_t,n)}$ as
\begin{subequations}
	\begin{align}\hspace{-0.4cm}
		&\underset{\substack{\mtW_{u,n},\mtM_{u,n},\\
				\hat{r}_{u,n},\mtE_{u,n}}}{\text{maximize}} \quad \sum\limits_{u=1}^U \beta_{u}\sum\limits_{n=1}^N    \hat{r}_{u,n}  \label{EQ:PROB_SCA_OBJ}&\\
		&\mbox{subject to \quad \eqref{PROB:MAX_RATE_LATENCY_CONS1}, \eqref{EQ:PROB_MSE_CONS3}, \eqref{EQ:PROB_MSE_CONS4}, \eqref{EQ:PROB_MSE_CONS5} and \eqref{EQ:TAYLOR}}.
	\end{align}
	\label{EQ:PROB_SCA}
\end{subequations}}
	Now, problem \eqref{EQ:PROB_SCA} is convex for either {$\{\mtW_{u,n}\}_{\forall(u\in\stU_t,n)}$ or  $\{\mtM_{u,n}\}_{\forall(u\in\stU_t,n)}$} when keeping the other variables fixed. 
Consequently, the beamforming design is an iterative process where the receive and transmit beamformers are alternately updated. The complete \ac{SCA} algorithm can be seen in Algorithm \ref{ALG:SCA_ALGORITHM}. 

We remark that the global optimality of the solution achieved by Algorithm~\ref{ALG:SCA_ALGORITHM} cannot be guaranteed, which occurs due to the iterative linear approximation procedure employed by the \ac{SCA} method~\cite{Boyd2007,Marks1978}. {In other words, unfortunately, we were not able to find the optimal solution of the non-trivial and non-convex problem \eqref{PROB:MAX_RATE_LATENCY}. However, as it is shown in Appendix~A, Algorithm~\ref{ALG:SCA_ALGORITHM} converges to a \ac{KKT} point of problem \eqref{PROB:MAX_RATE_LATENCY_REF}. Thus, we are capable of computing local optimal solutions to problem \eqref{PROB:MAX_RATE_LATENCY_REF}, which are attractive options considering the difficulty to solve this problem}. Furthermore, in the initialization phase of Algorithm~\ref{ALG:SCA_ALGORITHM}, one needs to obtain arbitrary feasible transmit beamformers\footnote{Weighted common rate maximization such as in \cite{Antonioli2019} can be used to obtain a feasible and valid initial point for Algorithm~\ref{ALG:SCA_ALGORITHM}.} {$\{\mtM^{(0)}_{u,n}\}_{\forall(u\in\stU_t,n)}$} so that the transmit power constraint and rate constraints of problem \eqref{EQ:PROB_SCA} are satisfied. Once problem \eqref{EQ:PROB_SCA} has been solved, the current \ac{MSE} values {$\{\mtE^{(k)}_{u,n}\}_{\forall(u\in\stU_t,n)}$} are used to update the point of approximation for the next iteration, {$\mtE_{u,n}^{(k+1)}$}, so that constraints \eqref{EQ:TAYLOR} hold with equality $\forall(u\in\stU_t,n)$. 
	\begin{algorithm}[t!]
		\caption{Centralized algorithm using SCA.}\label{ALG:SCA_ALGORITHM}
		\begin{algorithmic}[1]
			\STATE Initialize {$\{\mtM^{(0)}_{u,n}, \mtE_{u,s,n}^{(0)}\}_{\forall(u\in\stU_t,n)}$}.
			\REPEAT
			\STATE Generate {$\{\mtW_{u,n}\}_{\forall(u\in\stU_t,n)}$} using~\eqref{EQ:MMSE}.
			\STATE Set $k = 0$.
			\REPEAT
			\STATE Solve {$\{\mtM_{u,n}\}_{\forall (u\in\stU_t,n)}$} from~\eqref{EQ:PROB_SCA}.
			\STATE Set $k = k + 1$.
			\STATE Update {$\{\mtE_{u,n}^{(k)}\}_{\forall(u\in\stU_t,n)}$} from~\eqref{EQ:PROB_SCA}.
			\UNTIL{Convergence has been reached or $k > I_{max}$.}
			\UNTIL{Convergence has been reached.}
		\end{algorithmic}
	\end{algorithm}
	
	Algorithm \ref{ALG:SCA_ALGORITHM} is executed by a central controlling unit, which is responsible for computing all the transmit and receive beamformers using global \ac{CSI}. Multiple \ac{SCA} updates can be performed for each fixed receive beamformer update until the convergence has been achieved or a maximum number of iterations, $I_{\max}$, has been performed. Upon convergence of the algorithm, the central controlling unit sends the optimized transmit beamformers to the respective \acp{BS} for data transmission, while linear \ac{MMSE} receivers are used for data reception.	
	
	\section{Decentralized Solution and Signaling Aspects}
	\label{SEC:DECENTRALIZED_SOLUTION}
	As we saw before, centralized solutions require global \ac{CSI} availability at a central controlling unit for performing the transmit and receive beamforming computations. {However, such central controlling units are not always available, in which case distributed solutions are desirable. This can be the case in mobile networks, in which powerful centralized computational entities are not (yet) deployed in a cloud node, or when the mobile network operator prefers to
	deploy decentralized computations in the radio access network and fiber optical networks connecting the base stations.} 
	Therefore, in this section, we propose a decentralized solution where the adaptation of variables is executed distributedly among the nodes (users and \acp{BS}). In addition, to address the need for exchanging information between nodes in decentralized solutions, 
we present a signaling scheme in Section \ref{SEC:SIGNALING} to enable the decentralized processing.
	
	\subsection{Decentralized solution}
	Due to the interference terms and rate constraints present in the transmit beamformer update phase, 
the optimization problem \eqref{EQ:PROB_MSE} is, in general, not decoupled among \acp{BS}.
	Therefore, we propose a decentralized solution by initially applying a partial Lagrangian relaxation of the rate constraints \eqref{EQ:PROB_MSE_CONS3} and \eqref{EQ:PROB_MSE_CONS4}. The proposed relaxed formulation of problem \eqref{EQ:PROB_MSE} is given as:
	{\makeatletter 
	\if@twocolumn
		\begin{subequations}
		\begin{align}\hspace{-0.4cm}
			&\underset{\substack{\mtW_{u,n}, \mtM_{u,n},\\ \hat{r}_{u,n},\mtE_{u,n}}}{\text{maximize}} \quad  \sum\limits_{u=1}^U \beta_{u}\sum\limits_{n=1}^N \hat{r}_{u,n} - \sum\limits_{u=1}^{U}\gamma_{u}\Big(R_{u}^{\min} - \sum\limits_{n=1}^N \hat{r}_{u,n}\Big)\hspace{-20ex}&\nonumber \\
			&\hspace{12ex}- \sum\limits_{u=1}^{U}\phi_u\Big(\sum\limits_{n=1}^N \hat{r}_{u,n} - \frac{Q_u}{\Delta_\text{TTI}}\Big)\hspace{-20ex}& \\
			&\mbox{subject to \quad \eqref{PROB:MAX_RATE_LATENCY_CONS1}, \eqref{EQ:PROB_MSE_CONS2} and \eqref{EQ:PROB_MSE_CONS5}.}
		\end{align}
		\label{EQ:PROB_MSE_LAG_REL}
	\end{subequations}
\else
		\begin{subequations}
		\begin{align}\hspace{-0.4cm}
			&\underset{\substack{\mtW_{u,n}, \mtM_{u,n},\\ \hat{r}_{u,n},\mtE_{u,n}}}{\text{maximize}}  \sum\limits_{u=1}^U \beta_{u}\sum\limits_{n=1}^N \hat{r}_{u,n} - \sum\limits_{u=1}^{U}\gamma_{u}\Big(R_{u}^{\min} - \sum\limits_{n=1}^N \hat{r}_{u,n}\Big) - \sum\limits_{u=1}^{U}\phi_u\Big(\sum\limits_{n=1}^N \hat{r}_{u,n} - \frac{Q_u}{\Delta_\text{TTI}}\Big)\hspace{-20ex}& \\
			&\mbox{subject to \eqref{PROB:MAX_RATE_LATENCY_CONS1}, \eqref{EQ:PROB_MSE_CONS2} and \eqref{EQ:PROB_MSE_CONS5}.}
		\end{align}
		\label{EQ:PROB_MSE_LAG_REL}
	\end{subequations}
\fi \makeatother}
	However, due to constraints \eqref{EQ:PROB_MSE_CONS2} and \eqref{EQ:PROB_MSE_CONS5},
this reformulation is still coupled among \acp{BS}. Then, we apply the primal-dual method~\cite{Bertsekas1999} aiming {to solve} \eqref{EQ:PROB_MSE_LAG_REL} in a decentralized way, where the dual variables $\{\gamma_{u},\ \phi_u\}_{\forall u\in\stU_t}$ are fixed while solving the primal problem \eqref{EQ:PROB_MSE_LAG_REL} and updated according to the violation of the corresponding constraints.
	
The primal problem \eqref{EQ:PROB_MSE_LAG_REL} is solved iteratively. Thus, we begin by fixing the receive beamforming vectors to be the \ac{MMSE} receive beamformers \eqref{EQ:MMSE} and then apply the convex approximation in constraints \eqref{EQ:PROB_MSE_CONS2}, obtaining:
	{\makeatletter 
	\if@twocolumn
	\begin{subequations}
	\begin{align}\hspace{-0.4cm}
		&\underset{\substack{\mtM_{u,n}, \hat{r}_{u,n},\\\mtE_{u,n}}}{\text{maximize}} \quad \sum\limits_{u=1}^U \beta_{u} \sum\limits_{n=1}^N \hat{r}_{u,n} - \sum\limits_{u=1}^{U}\gamma_{u}\Big(R_{u}^{\min} - \sum\limits_{n=1}^N\hat{r}_{u,n}\Big)\hspace{-20ex}&\nonumber \\
		&\hspace{12ex}- \sum\limits_{u=1}^{U}\phi_u\Big(\sum\limits_{n=1}^N\hat{r}_{u,n} - \frac{Q_u}{\Delta_\text{TTI}}\Big)\hspace{-20ex}& \\
		&\mbox{subject to \quad \eqref{PROB:MAX_RATE_LATENCY_CONS1}, \eqref{EQ:PROB_MSE_CONS5} and \eqref{EQ:TAYLOR}.}
	\end{align}
	\label{EQ:PROB_SCA_LAG_REL}
\end{subequations}
	\else
			\begin{subequations}
			\begin{align}\hspace{-0.4cm}
				&\underset{\substack{\mtM_{u,n}, \hat{r}_{u,n},\\\mtE_{u,n}}}{\text{maximize}}  \sum\limits_{u=1}^U \beta_{u} \sum\limits_{n=1}^N \hat{r}_{u,n} - \sum\limits_{u=1}^{U}\gamma_{u}\Big(R_{u}^{\min} - \sum\limits_{n=1}^N\hat{r}_{u,n}\Big) - \sum\limits_{u=1}^{U}\phi_u\Big(\sum\limits_{n=1}^N\hat{r}_{u,n} - \frac{Q_u}{\Delta_\text{TTI}}\Big)\hspace{-20ex}& \\
				&\mbox{subject to \eqref{PROB:MAX_RATE_LATENCY_CONS1}, \eqref{EQ:PROB_MSE_CONS5} and \eqref{EQ:TAYLOR}.}
			\end{align}
			\label{EQ:PROB_SCA_LAG_REL}
		\end{subequations}
	\fi \makeatother}
	
	We solve the \ac{KKT} conditions of problem \eqref{EQ:PROB_SCA_LAG_REL} by assuming that \eqref{EQ:PROB_MSE_CONS5} and \eqref{EQ:TAYLOR} are tight. Thus, from the \ac{KKT} conditions of problem \eqref{EQ:PROB_SCA_LAG_REL}, the dual variables, {$\{\psi_{u,n}\}_{\forall(u\in\stU_t,n)}$}, related to constraints \eqref{EQ:TAYLOR} are computed as:
	\begin{equation}
	\psi_{u,n} = [\beta_{u} + \gamma_{u} - \phi_{u}]^{+}. \label{EQ:PSI}
	\end{equation}
	Meanwhile, the  dual variables related to constraints \eqref{EQ:PROB_MSE_CONS5}, denoted as {$\bm{\theta}_{u,n}\in\bbC^{S\times S}\  {\forall(u\in\stU_t,n)}$}, are updated as follows:
	{\begin{align}
		\bm{\theta}_{u,n}^{(k+1)} = \bm{\theta}_{u,n}^{(k)} +  \rho^{(k)}\Bigg(\frac{\psi_{u,n}}{\mtE_{u,n}^{(k)}\log(2)}  - \bm{\theta}_{u,n}^{(k)}\Bigg), \label{EQ:THETA}
	\end{align}
	where each element of $\bm{\theta}_{u,n}^{(k+1)}$ can be interpreted as a point in the line segment between each element in $\bm{\theta}_{u,n}^{(k)}$ and $\frac{\psi_{u,n}}{\mtE_{u,n}^{(k)}\log(2)}$ 
	determined by a diminishing or fixed step size $\rho^{(k)}\in(0,1)$.} The choice of $\rho^{(k)}$ is system-dependent and
	its value affects the convergence behavior and also controls
	the oscillations in the users’ rate when \eqref{EQ:PSI} is negative (before projection) due to over-allocation of resources. In other words, when the  achievable rate of a given user is greater than the amount of bits available in its buffer, \eqref{EQ:PSI} can be zero, consequently, {$\bm{\theta}^{(k+1)}$ is element-wise lower than $\bm{\theta}^{(k)}$}, as seen in \eqref{EQ:THETA}. As we will see later, the dual variable {$\bm{\theta}_{u,n}^{(k)}$} acts as precoder weight for computing {$\mtM_{u,n}$}, thus, when reducing {the elements of $\bm{\theta}_{u,n}^{(k)}$}, the achievable rate decreases in order to avoid over-allocation of resources.
	
	From the \ac{KKT} conditions of \eqref{EQ:PROB_SCA_LAG_REL}, we can solve the transmit beamformers, $\{\mtM_{u,n}\}_{\forall (u\in\stU_t,n)}$, as follows:
	{\begin{equation}
	\mtM_{u,n} = (\mtQ_{b_u} + \nu_{b_u}\mtI)^{-1}\Herm{\mtH}_{b_{u},u,n}\mtW_{u,n}\bm{\theta}_{u,n}, \label{EQ:TRANSMIT_BEAMFORMERS}
\end{equation}
	where $\mtQ_{b_u} = \sum_{i=1}^{U} \Herm{\mtH}_{b_{u},i,n}\mtW_{i,n}\bm{\theta}_{i,n}\Herm{\mtW}_{i,n}\mtH_{b_{u},i,n}$} and $\nu_{b_u}$ is the dual variable associated to the power budget constraints of \eqref{EQ:PROB_SCA_LAG_REL}. From \eqref{EQ:TRANSMIT_BEAMFORMERS} we can observe that {$\{\bm{\theta}_{u,n}\}_{\forall (u,n)}$
	act as weights of user $u$ on sub-channel $n$}. The value of $\nu_{b_u}\geq 0$ should be chosen to meet the complementary
	slackness condition of the power budget constraints. Note that if the power constraint is not active when solving \eqref{EQ:TRANSMIT_BEAMFORMERS} for $\nu_{b_u} = 0$, then
	the beamformers are optimal. Otherwise, the optimal value of
	$\nu_{b_u}$ can be obtained using one dimensional search  techniques (e.g., bisection method) with respect to the power budget constraints \cite{Shi2011}. The high complexity due to the matrix inversion in \eqref{EQ:TRANSMIT_BEAMFORMERS}
	can be reduced by using an eigenvalue decomposition of $\mtQ_{b_u} + \nu_{b_u}\mtI$, as shown in \cite{Shi2011}, or by solving the linear system {$(\mtQ_{b_u} + \nu_{b_u}\mtI)\mtM_{u,n} = \Herm{\mtH}_{b_{u},u,n}\mtW_{u,n}\bm{\theta}_{u,n}, \forall (u\in\stU_t,n)$.}
	
	Once the current \ac{MSE} values, {$\{\mtE_{u,n}\}_{\forall(u\in\stU_t,n)}$} are computed, we update the variable $\hat{r}_{u,s,n}^{(k+1)}$ as:
	{\begin{equation}
		\hat{r}_{u,n}^{(k+1)} = -\log_2\det\left(\mtE_{u,n}^{(k)}\right) - \frac{\Trace{\left(\mtE_{u,n}^{(k)}\right)^{-1}\left(\mtE_{u,n} - \mtE_{u,n}^{(k)}\right)}}{\log(2)}.\label{EQ:R}
	\end{equation}}
	In addition, the \ac{SCA} operating point is also updated with the current \ac{MSE} value, i.e., {$\mtE_{u,n}^{(k+1)} = \mtE_{u,n}$, $\forall(u\in\stU_t,n)$}. Finally, in the dual update, the rate demand weight factors $\{\gamma_{u}\}_{\forall u\in\stU_t}$ and queue weight factors
	$\{\phi_u\}_{\forall u\in\stU_t}$ follow, from their respective constraint violations, as
	\begin{equation}
		\gamma_{u}^{(k+1)} = \Bigg(\gamma_{u}^{(k)} + \rho^{(k)}\Big(R_u^{\min} - {\sum_{n=1}^{N}\hat{r}_{u,n}^{(k+1)}}\Big)\Bigg)^{+}, \label{EQ:GAMMA}
	\end{equation}
	and
	\begin{equation}
		\phi_{u}^{(k+1)} = \Bigg(\phi_{u}^{(k)} + \rho^{(k)}\Big({\sum_{n=1}^{N}\hat{r}_{u,n}^{(k+1)}} - \frac{Q_u}{\Delta_\text{TTI}}\Big)\Bigg)^{+}. \label{EQ:PHI}
	\end{equation}
	This also corresponds to a subgradient update of the dual variables in terms of \eqref{EQ:PROB_SCA_LAG_REL} with the approximated rate constraints, where setting an appropriate value for the step size plays an important role (for more details, see \cite{Bertsekas1999, Palomar2006}).
	
	\begin{algorithm}[!t]
		\caption{Decentralized algorithm.}\label{ALG:DEC_ALG_SCA}
		\begin{algorithmic}[1]
			\STATE Initialize {$\{\mtM^{(0)}_{u,n}, \mtE_{u,n}^{(0)}, \bm{\theta}_{u,n}^{(0)}\}_{\forall (u\in\stU_t,n)}$} and $\{\gamma_{u}^{(0)}, \phi_{u}^{(0)}\}_{\forall u\in\stU_t, n}$.
			\STATE \textit{BS}: Use initial {$\{\mtM_{u,n}\}_{\forall (u\in\stU_t,n)}$} to transmit pilots.	
			\REPEAT		
			\STATE \textit{UE}: Generate {$\mtW_{u,n}$} using~\eqref{EQ:MMSE}.
			\STATE Set $k = 0$.
			\REPEAT
			\STATE \textit{UE}: Measure \ac{MSE} {$\mtE_{u,n}$} as shown in~\eqref{EQ:MSE}.
			\STATE \textit{UE}: Compute ${\hat{r}_{u,n}}$ using \eqref{EQ:R}.
			\STATE \textit{UE}: Update variable ${\psi_{u,n}}$ from~\eqref{EQ:PSI}.
			\STATE \textit{UE}: Update variable $\gamma_{u}$ from~\eqref{EQ:GAMMA}.
			\STATE \textit{UE}: Update variable $\phi_{u}$ from~\eqref{EQ:PHI}.
			\STATE \textit{UE}: Update the weights $\bm{\theta}_{u,n}$ from~\eqref{EQ:THETA}.
			\STATE \textit{UE}: Send $\bm{\theta}_{u,n}$ to \ac{BS} using uplink signaling.
			\STATE \textit{BS}: Exchange {$\{\bm{\theta}_{u,n}\}_{\forall (u\in\stU_t,n)}$} via backhaul link.
			\STATE \textit{BS}: Solve {$\{\mtM_{u,n}\}_{\forall (u\in\stU_t,n)}$} from~\eqref{EQ:TRANSMIT_BEAMFORMERS}.
			\STATE \textit{BS}: Use {$\{\mtM_{u,n}\}_{\forall (u\in\stU_t,n)}$} to transmit pilots.	
			\STATE \textit{UE}: {$\mtE_{u,n}^{(k+1)}$ $\leftarrow$ $\mtE_{u,n}$}.		
			\STATE Set $k = k + 1$.
			\UNTIL{Convergence has been reached or $k > I_{max}$.}
			\UNTIL{Convergence has been reached.}
		\end{algorithmic}
	\end{algorithm}
		
	Algorithm~\ref{ALG:DEC_ALG_SCA} describes the proposed decentralized solution.
	As we can see, multiple consecutive \ac{SCA} updates can be performed for each fixed receive beamformer update, such as in Algorithm~\ref{ALG:SCA_ALGORITHM}. It is worth mentioning that the weights {$\{\bm{\theta}_{u,n}\}_{\forall (u\in\stU_t,n)}$} depend only on the instantaneous \ac{MSE} values {$\{\mtE_{u,n}\}_{\forall (u\in\stU_t,n)}$}, while the variables $\{\gamma_{u},\phi_{u}\}_{\forall u\in\stU_t}$ are computed using only the current rate value of user $u$. Therefore, assuming the knowledge of the received signal covariance, these variables can be computed locally at each user. Then, using some signaling strategy (which is discussed later), such information can be transmitted to the \acp{BS}.
	The convergence analysis of Algorithm~\ref{ALG:DEC_ALG_SCA} is shown in Appendix~A.

	To the best of our knowledge, methods for finding feasible initialization points, such as in \cite{Antonioli2019}, require centralized processing, which can be critical for decentralized solutions. {However,
	differently from the centralized solution, the rate constraints are not required to be feasible at each iteration of the proposed decentralized algorithm. Therefore, it is not necessary to find a feasible initialization point \cite{Kaleva2016}.
	This is accomplished by means of the non-trivial partial Lagrangian relaxation followed by a dual-based decomposition that we apply when developing the proposed decentralized solution.}
		
	In general, the algorithm will find a feasible solution, mainly, when the power budget and maximum rate requirements are large in comparison to the minimum rate requirements. On the other hand, the algorithm can fail to find a feasible solution when the power budget is (severely) tight and the feasible regions around the locally optimal points are restricted. See \cite{Bertsekas1999} for more details about ill-conditioned problem formulation. Observe that, even if the algorithm fails in finding a feasible solution, the algorithm can still find a region where the rate constraints violation is reasonably small. Indeed, for non-feasible rate
	constraints, the rate demand variables, $\gamma_{u}$, will increase until the minimum rate constraints are satisfied. The same occurs for queue weight variables, $\phi_{u}$, which increase until the maximum rate constraints are fulfilled.
		
\subsection{Signaling Aspects}
\label{SEC:SIGNALING}
In this section, we propose a signaling framework 
for practical implementation of the proposed decentralized algorithm, which uses precoded pilots and relies on backhaul signaling.
	
The signaling scheme adopted herein is based on \cite{Tolli2019}, 
and extended to our proposed framework. Thus, only local \ac{CSI} is required to be available at the \acp{BS} and users. The proposed precoded pilot scheme is required for the update of the receive beamforming {matrices} 
from \eqref{EQ:MMSE} at the user side and transmit beamforming {matrices} from \eqref{EQ:TRANSMIT_BEAMFORMERS} 
at the \acp{BS}, where both \eqref{EQ:MMSE} and \eqref{EQ:TRANSMIT_BEAMFORMERS} require some knowledge about interfering \ac{CSI}.
	
Considering the acquisition of \ac{CSI} needed to compute the \ac{MMSE} receivers in \eqref{EQ:MMSE} at the user side, besides the local channel from \ac{BS} $b_u$ to user $u$, the channel information that needs to be acquired by user $u$ is $\mtH_{b_{i},u,n}$, which is the channel between user $u$ and all the \acp{BS} $b_i$ in the system. User $u$ does not need to have knowledge of the channel from a given \ac{BS} $b_i$ and any other user $\tilde{u} \neq u$.  Nevertheless, acquiring the channel matrix $\mtH_{b_{i},u,n}$ in itself is a very difficult task. To this end,
what user~$u$ can actually estimate is the effective channel {$\mtH_{b_{i},u,n}\mtM_{i,n}$}, 
which accounts for the interference caused by all \acp{BS} in the system when transmitting to all the users in all streams for sub-channel~$n$. 
In fact, the effective channels {$\mtH_{b_{i},u,n}\mtM_{i,n}$, for $i=1,\ldots,U$} can be estimated by user $u$ using the precoded pilot signaling scheme adopted herein.  Therefore, in the proposed scheme all nodes use known orthogonal precoded pilot symbols, allowing perfect signal separation and estimation of the effective channels. Specifically, users should be more aware of the neighborhood and measure the base stations in the near vicinity, i.e., the users should be able to measure pilots from the \acp{BS} in the system in order to be able to compute their respective \ac{MMSE} receive beamformers. More details can also be found in~\cite{Tolli2019}.

Considering the acquisition of \ac{CSI} required to compute the transmit beamforming vectors in \eqref{EQ:TRANSMIT_BEAMFORMERS} at the \ac{BS} side, each \ac{BS} requires the knowledge of the effective channels from all users in the system to itself. Similarly to what is done at the user side, the interfering \ac{CSI} can be obtained by each base station by measuring the precoded pilots sent by all users in the system. Consequently, the same assumptions described above all hold for the \ac{CSI} estimation at the \ac{BS}.
		
\begin{figure}[!t]
		\centering
		\includegraphics[width=0.5\columnwidth]{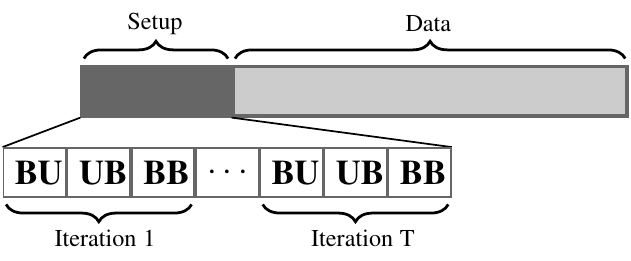}
		\caption{Frame structure.}
		\label{FIG:FRAME_STRUCTURE}
	\end{figure}

During the execution of Algorithm \ref{ALG:DEC_ALG_SCA}, the precoded pilots are transmitted by means of an over-the-air signaling between users and \acp{BS}, while backhaul signaling is required for communications between \acp{BS}. Given these considerations, \FigRef{FIG:FRAME_STRUCTURE} presents the proposed frame structure, which is a modified version of the frame structure proposed in \cite{Tolli2019}.
	
\begin{figure}[!t]
		\centering
		\vspace{-3ex}
		\includegraphics[width=0.5\columnwidth]{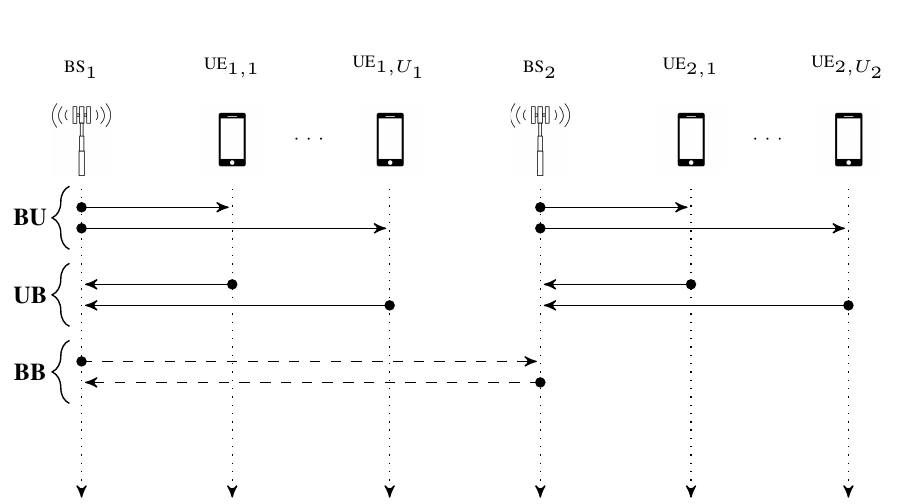}
		\caption{Signaling exchange illustration in a multicell MIMO-OFDM scenario.}
		\label{FIG:SIGNALING}
	\end{figure}

	The frame structure in~\FigRef{FIG:FRAME_STRUCTURE} is split into two parts: a beamformers setup phase and data transmission phase. The over-the-air and backhaul signaling occur during the beamformers setup phase, where the over-the-air signaling is divided into two phases, more precisely, the forward precoded pilot transmission from \ac{BS} to the users, denoted as \textbf{BU}, which occurs in lines 2 and 16 of Algorithm \ref{ALG:DEC_ALG_SCA}, and the backward signaling from each user to its serving \ac{BS}, namely \textbf{UB}, which occurs in line 13 of Algorithm \ref{ALG:DEC_ALG_SCA}.
	Finally, the backhaul signaling is used to share the weights {$\{\bm{\theta}_{u,n}\}_{\forall(u\in\stU_t,n)}$} between \acp{BS} in line 14 of Algorithm \ref{ALG:DEC_ALG_SCA}, which is denoted as \textbf{BB}. The signaling exchange for a multicell \ac{MU-MIMO} with two \acp{BS} is illustrated in~\FigRef{FIG:SIGNALING}.

\subsection{Computational Complexity and Signaling Overhead}
	
The computational complexity of the decentralized algorithm (i.e., Algorithm 2) is dominated by the matrix inversion in \eqref{EQ:MMSE} and \eqref{EQ:TRANSMIT_BEAMFORMERS}, and the \ac{MSE} computation in \eqref{EQ:MSE}. It should be noted that \eqref{EQ:MMSE}, \eqref{EQ:MSE} and \eqref{EQ:TRANSMIT_BEAMFORMERS} are also required by the conventional WMMSE approaches in \cite{Shi2011, Kaleva2016, Venkatraman2016}, which correspond to the baseline algorithms considered in this paper. Therefore, the per-iteration and per-subchannel computational complexity of \eqref{EQ:MMSE}, \eqref{EQ:MSE} and \eqref{EQ:TRANSMIT_BEAMFORMERS} is in the order $\stO(U^2N_TN_R^2 + U^2N_T^2N_R + U^2N_T^3 + UN_R^3)$. Therefore, assuming that the computation of other variables can be solved by linear expressions, whose contribution to the overall complexity can be ignored, the per-iteration and per-subchannel computational complexity of the proposed solution is also in the order $\stO(U^2N_TN_R^2 + U^2N_T^2N_R + U^2N_T^3 + UN_R^3)$. Such a per-iteration computational complexity can be easily handled by current base stations and user devices for a moderate number of transmit/receive antennas.
	
In addition, the proposed solution can be implemented in a distributed fashion by applying a precoded pilot signaling scheme, which is described in Section~\ref{SEC:SIGNALING}. In that signaling scheme, each iteration has an associated overhead due to the transmission of precoded uplink/downlink pilots. Based on \cite{Tolli2019}, we can measure the communication overhead by the number of orthogonal pilot symbols needed for each iteration, which is given by $\Omega = 2TBU_bS$, where $T$ is the number of iterations. Thus, the minimum number of orthogonal pilots, $\Omega$, increases with the number of data streams, base stations, users, and iterations. Therefore, increasing the number of inner iterations of Algorithm~2 incurs a higher signaling overload.  Thus, in order to obtain a practical implementation of Algorithm~2 with minimal signaling overhead, this number of iterations can be limited to a maximum of 10 iterations per data frame, as suggested in~\cite{Tolli2019}, at the cost of a possibly lower performance in some situations. Under these conditions, the proposed decentralized algorithm has the potential to handle moderate latency-sensitive applications.
	
Considering the above discussion, we conclude that the proposed solution has a computational complexity comparable with that of existing solutions, which can be handled by existing hardware, and can be efficiently deployed based on the proposed precoded pilot scheme.\\
					
\section{Performance Evaluation}
\label{SEC:PERFORMANCE_EVALUATION}
	
In this section we present the simulation setting and simulation results. More precisely, Section~\ref{SEC:SIMULATION_ASSUMPTIONS} details the simulation setup. The convergence analysis of the proposed solutions is conducted in Section~\ref{SEC:CONVERGENCE_ANALYSIS}, while the performance evaluation using a Poisson traffic model is presented in Section~\ref{SEC:POISSON_TRAFFIC_MODEL}. Section~\ref{SEC:BURSTY_TRAFFIC_MODEL} shows the performance of the proposed solution under a bursty traffic model and, finally, we analyze the impact of imperfect \ac{CSI} in Section~\ref{SEC:IMPERFECT_CSI}.
	
\subsection{Simulation Assumptions}
\label{SEC:SIMULATION_ASSUMPTIONS}
We consider the downlink of multicell \ac{MU-MIMO} \ac{OFDM} scenarios where each \ac{BS} is located at the center of a hexagonal cell and the users are uniformly distributed within the cell. We set the inter-site distance to 250 m. Moreover, uniform linear arrays are employed by all \acp{UE} and \acp{BS}. The \ac{BS} and \acp{UE} heights are 25~m and 1.5~m, respectively, and the \acp{UE} speed is equal to 3 km/h. The \ac{5G-StoRM} \cite{Pessoa2019,3GPP2018Study}, assuming the \ac{UMi} scenario, is used for all links, considering a carrier frequency of 2 GHz and that each sub-channel has a bandwidth of 180 kHz. {More details about the channel generation can be found in \cite{Pessoa2019,3GPP2018Study}.} For all analyzed scenarios, the power budget is $P_b$ = 35 dBm, $\forall b\in\stB$ and the step size, $\rho^{(k)}$, is fixed to be equal to 0.01. 
	Also, every user has the same average packet arrival rate and packet size in the simulation. Unless otherwise stated, $\bar{L}_{u,m} = 9600$ bits. Based on \cite{3GPP2018Service}, the maximum outage probability and the maximum tolerable latency are set as 0.05 and 20 ms, respectively (i.e., $\xi = 0.05$ and $d_{\max} = 20$ ms).
	The simulations are performed during 300 time slots or \acp{TTI}, where each \ac{TTI} has a duration of 1 ms. Also, the results are obtained from 100 Monte-Carlo simulations.
	
	We consider three state-of-the-art solutions for performance comparison against our proposed solution.
	The first solution is the \ac{WMMSE} algorithm~\cite{Shi2011}, which solves a weighted sum-rate maximization problem without \ac{QoS} constraints. The second solution is the \ac{JSFRA} algorithm~\cite{Venkatraman2016}, which minimizes the total number of backlogged packets in each \ac{TTI}.
	Finally, we also consider the algorithm proposed in~\cite{Kaleva2016}, which solves the weighted sum-rate maximization problem with minimum rate requirements (hereafter, the solution from~\cite{Kaleva2016} is referred to as Kaleva).
	Note that both \ac{WMMSE} and Kaleva algorithms were conceived assuming a full-buffer model, while the \ac{JSFRA} solution considers a finite-buffer model.
	
\subsection{Convergence Analysis}
\label{SEC:CONVERGENCE_ANALYSIS}

\makeatletter 
\if@twocolumn
\begin{figure}[!t]
	\centering
	\includegraphics[width=0.8\columnwidth]{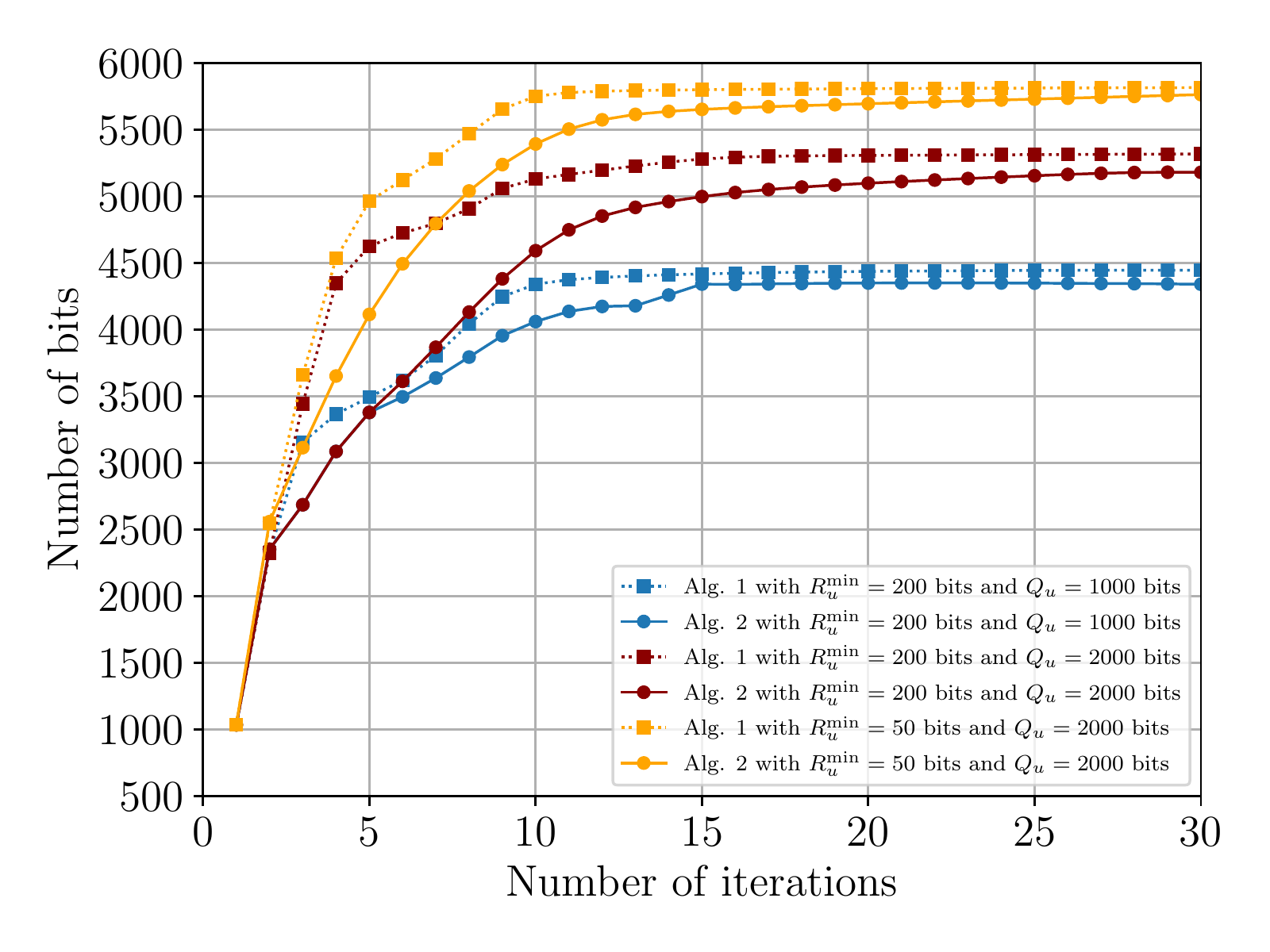}
	\caption{Convergence of the total number of transmitted bits for Algs. \ref{ALG:SCA_ALGORITHM} and \ref{ALG:DEC_ALG_SCA} with $\{B, U, U_b, N, N_\text{T}, N_\text{R}\} = \{2, 6, 3, 1, 8, 2\}$.}
	\vspace{-2ex}
	\label{FIG:CONVERGENCE_ALGORITHMS}
\end{figure}
\else
\begin{figure}[!t]
	\centering
	\includegraphics[width=0.5\columnwidth]{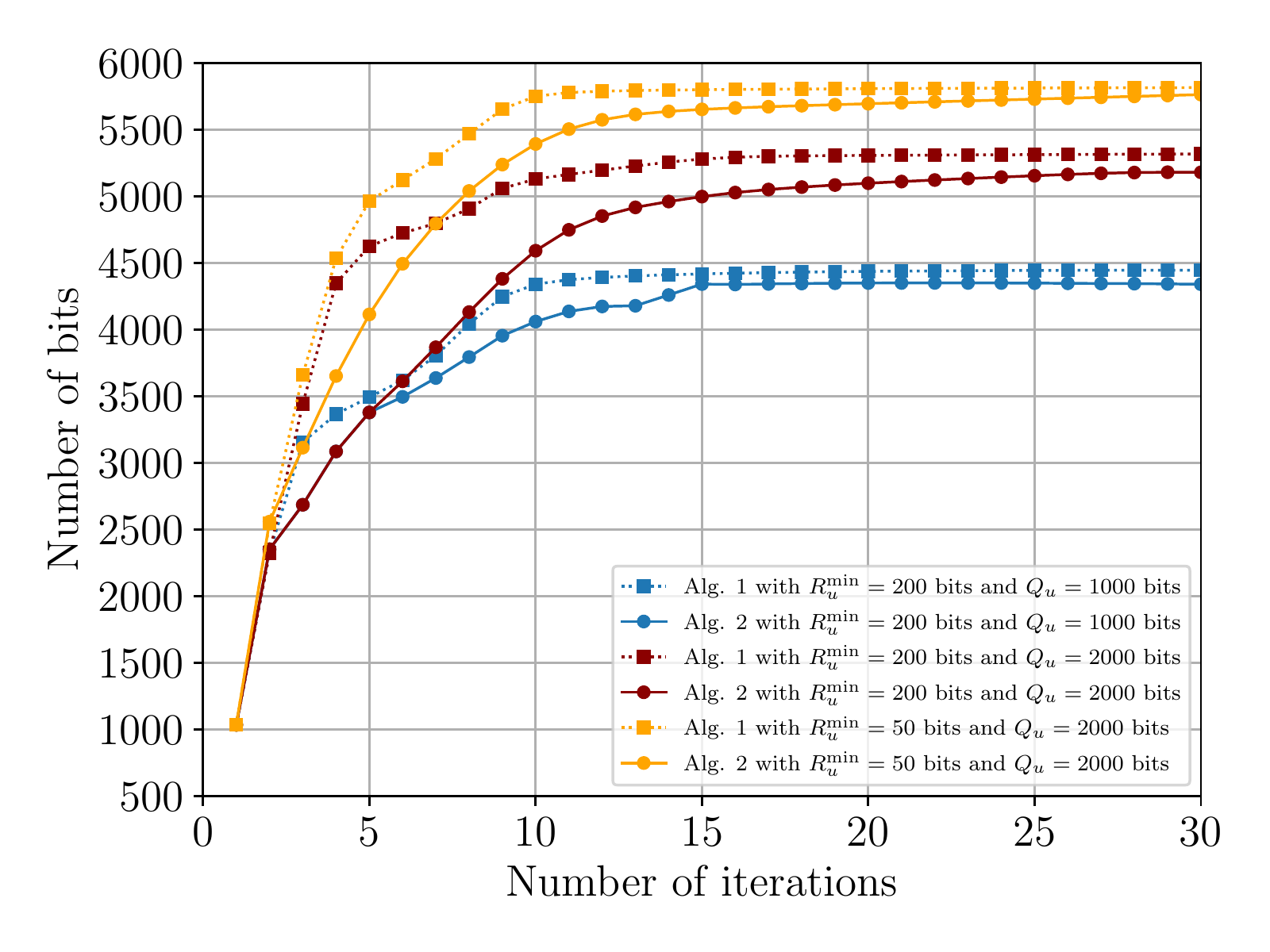}
	\caption{Convergence of the total number of transmitted bits for Algs. \ref{ALG:SCA_ALGORITHM} and \ref{ALG:DEC_ALG_SCA} with $\{B, U, U_b, N, N_\text{T}, N_\text{R}\} = \{2, 6, 3, 1, 8, 2\}$.}
	\vspace{-2ex}
	\label{FIG:CONVERGENCE_ALGORITHMS}
\end{figure}
\fi \makeatother

\FigRef{FIG:CONVERGENCE_ALGORITHMS} depicts the total number of transmitted bits as a function of the number of iterations for different values of $R_u^{\min}$ and $Q_u$.
The proposed solutions converges to the final solution with a low number of iterations.
Moreover, when $Q_u$ increases, for a fixed value of $R_u^{\min}$, the proposed solutions achieve a higher number of transmitted bits. In general, users with good channel conditions should diminish their data rate in order to avoid over-allocation of resources. However, when $Q_u$ increases, more bits are available in the buffers, thus, these users can transmit more data and, consequently, increase the system total data rate. On the other hand, when we increase the minimum number of transmitted bits required per user, $R_u^{\min}$, for a fixed value of $Q_u$, we observe that the number of transmitted bits achieved by the proposed algorithm diminishes.
The reason behind this is that the proposed solution allocates more resources to users in poor channel conditions in order to fulfill the minimum number of transmitted bits for all users, consequently, diminishing the total number of transmitted bits. Finally, comparing the performance of the proposed solutions, we can see that the proposed decentralized
solution (Algorithm 2) performs very close to the centralized solution (Algorithm 1) in terms of number of transmitted bits for different data rate requirements. However, we remark that the centralized solution requires an initial feasible point of problem \eqref{EQ:PROB_SCA} to start the iterations in Algorithm \ref{ALG:SCA_ALGORITHM}, which demands high computational complexity. On the other hand, the decentralized solution presents a better trade-off between performance and computational complexity. Consequently, we use only the decentralized solution for further performance evaluations.

\makeatletter 
\if@twocolumn
\begin{figure}[!t]
	\centering
	\includegraphics[width=0.8\columnwidth]{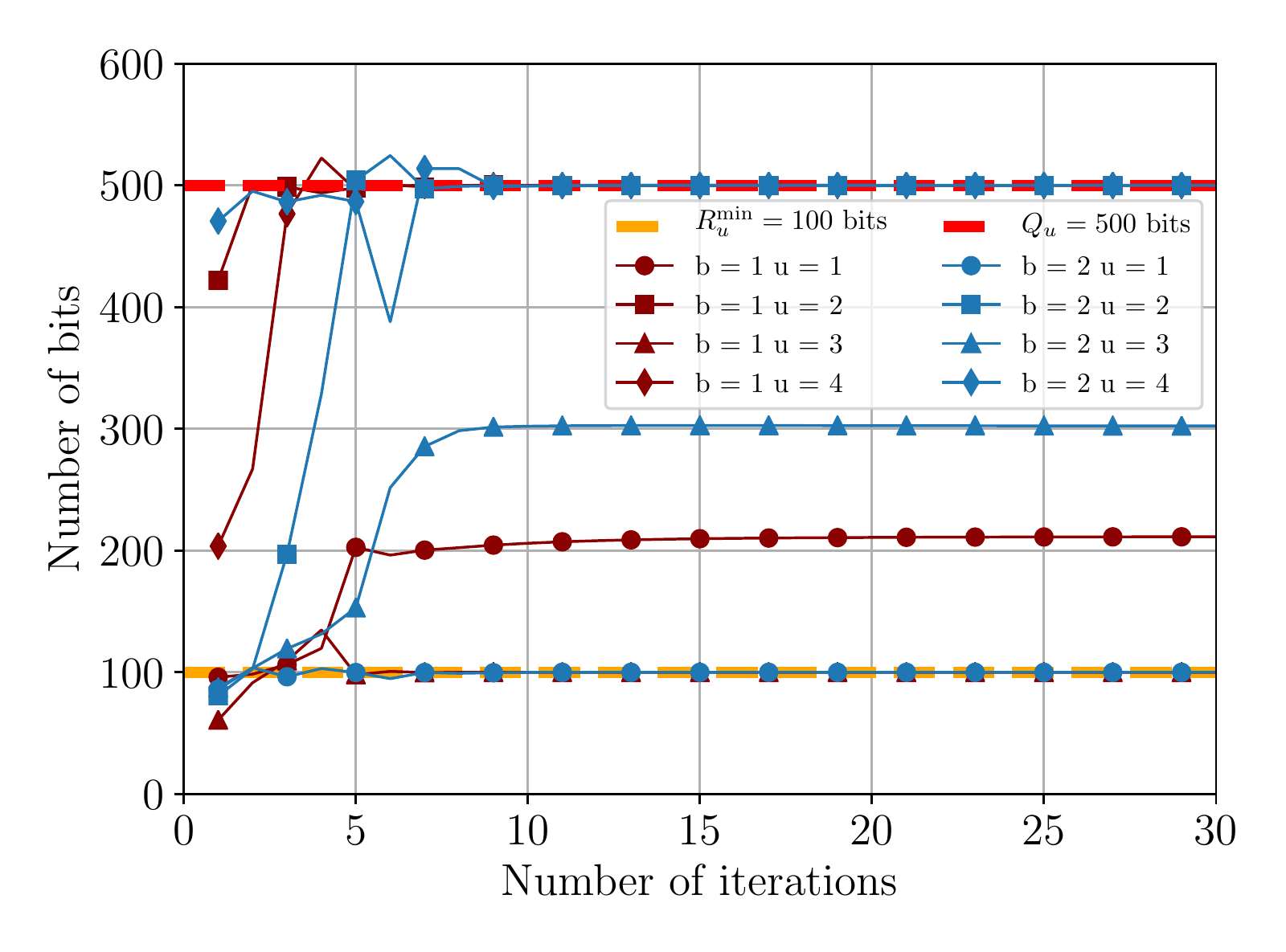}
	\caption{Convergence of the number of transmitted bits for all users with $\{B, U,U_b,N, N_T,N_R, R_u^{\min}, Q_u\} = \{2, 8, 4, 1, 4, 2, 100, 500\}$.}
	\label{FIG:CONVERGENCE_USERS}
\end{figure}
\else
\begin{figure}[!t]
	\centering
	\includegraphics[width=0.5\columnwidth]{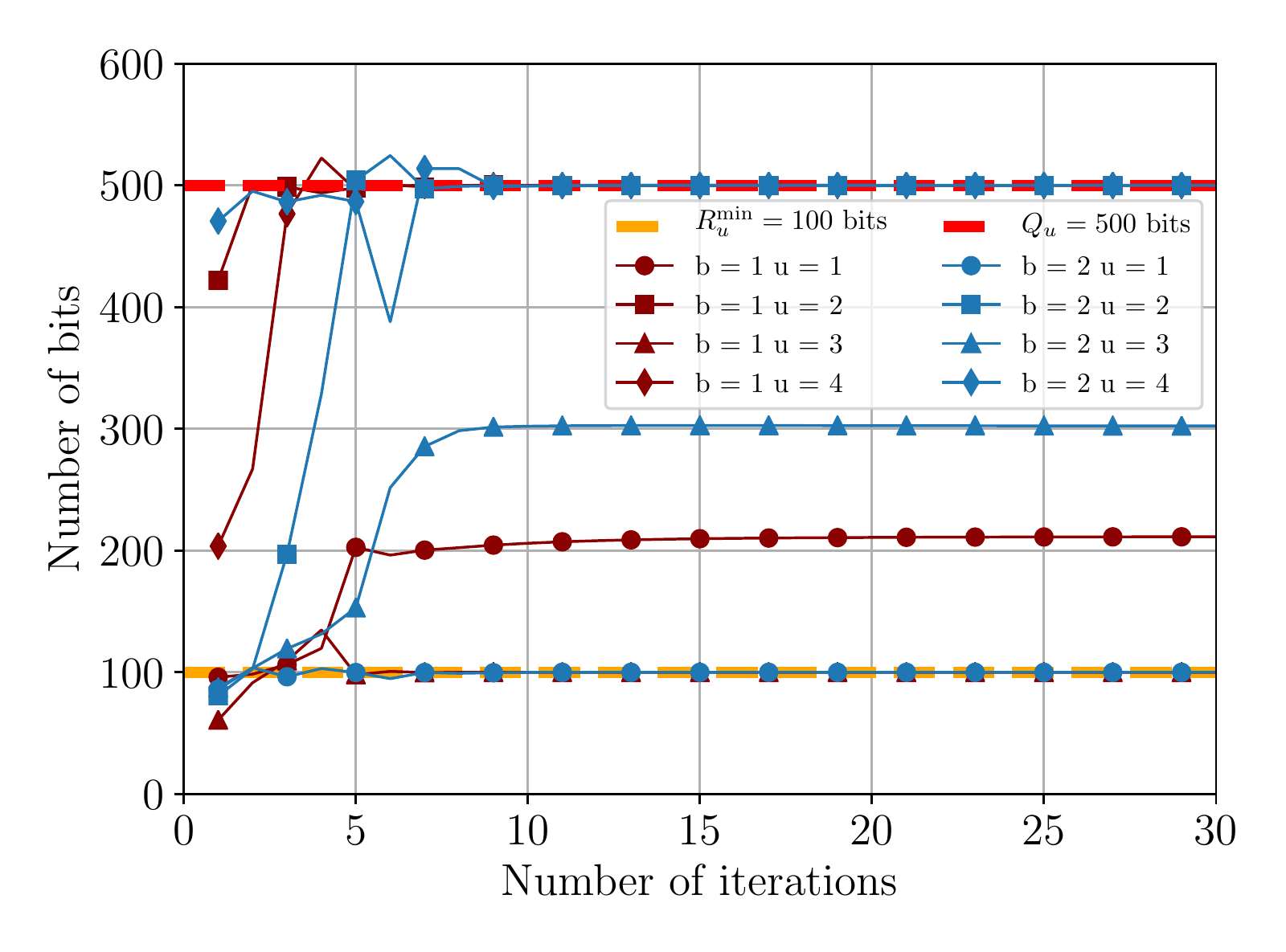}
	\caption{Convergence of the number of transmitted bits for all users with $\{B, U,U_b,N, N_T,N_R, R_u^{\min}, Q_u\} = \{2, 8, 4, 1, 4, 2, 100, 500\}$.}
	\label{FIG:CONVERGENCE_USERS}
\end{figure}
\fi \makeatother
In \FigRef{FIG:CONVERGENCE_USERS}, we show the convergence of the number of transmitted bits for all users for Algorithm \ref{ALG:DEC_ALG_SCA}. In the first iterations of the algorithm, only part of the users are assigned with a number of transmitted bits higher than the minimum requirement.
Nevertheless, as the algorithm converges, it adjusts the number of transmitted bits for the remaining users in order to fulfill the minimum requirement for all users. Moreover, it can be seen that some users, specially those with high channel gains, have the potential to transmit more bits than the number of available bits in their buffers. However, as the proposed solution converges, it reduces the amount of assigned transmitted bits for those users in order to avoid over-allocation of the resources. Finally, it is worth highlighting that only a low number of iterations is needed to achieve a good solution.
In the case illustrated in \FigRef{FIG:CONVERGENCE_USERS}, for example, 10 iterations would be enough to assure that all users are transmitting an amount of bits between $R_u^{\min}$ and $Q_u$.

\subsection{Poisson Traffic Model}
\label{SEC:POISSON_TRAFFIC_MODEL}
As previously mentioned, the Poisson traffic model still plays an important role in practice, as it is used for evaluation purposes in \ac{3GPP} analyses. Therefore, this subsection is concerned with a performance analysis considering a Poisson traffic model.

\FigRef{FIG:Outage} presents the outage probability versus the average arrival rate of packets, $\lambda_u$. For this analysis, we consider two different setups: (I) in the first setup, the user weights are set to 1 for all users, i.e., $\beta_{u} = 1,\ \forall u\in\stU_t$;  and (II) {in the second setup, four different user weights are assigned to the users}. Specifically, as we have 4 users per \ac{BS}, each user of a given \ac{BS} has a different weight. These user weights are kept fixed within each Monte-Carlo simulation, but change for different users in different Monte-Carlo simulations. {In this work, we assume two cases for the second setup: $\beta_{u} \in \{1,\ 3,\ 5,\ 7\}$ and $\beta_{u} \in \{1,\ 0.1,\ 0.01,\ 0.001\}$}

{First, }we can observe that the outage probability increases as the average packet arrival rate increases for all solutions. This occurs because more packets arrive at the users' buffers, leading to an increase in the waiting time of packets, consequently, increasing the number of outages.

{Also, note that different user weights have a significant impact on the \ac{WMMSE} performance. In fact, this is expected because the \ac{WMMSE} algorithm allocates more resources to users with high priority (higher user weights) in order to maximize the weighted sum-rate without considering the minimum data rate demands, leading to an increase in the outage probability. Regarding the \ac{JSFRA} algorithm, we observe an increase in the outage probability when $\beta_{u} = \{1, 0.1, 0.01, 0.001\}$. The reason for this is that, as the users' weights are not relatively close, the \ac{JSFRA} algorithm  tends to prioritize the users with high priority instead of the users with a higher number of queued packets. Consequently, the users with a higher number of queued packets tend to accumulate more bits in their buffers, affecting the outage probability performance of this solution. On the other hand, the outage probability of the Kaleva and the proposed solution remains almost unchanged for all setups\footnote{{It is worth noting that this is not necessarily the case for other metrics, such as the objective function in~\eqref{PROB:MAX_RATE_LATENCY}, which is clearly affected by the user weights. This means that the user weights have an impact on the computed  solution, so that we cannot always optimize the sum-rate problem instead of the weighted sum-rate problem.}}. This occurs because the average outage probability is related to minimum data rate requirements, which are not changed in both cases. Differently from the \ac{WMMSE} and \ac{JSFRA} algorithms, both the Kaleva and the proposed solutions take into account minimum QoS requirements. Consequently, even prioritizing the users with a higher priority, these solutions aim at meeting the \ac{QoS} constraints of each user, which causes a more fair distribution of the resources among users. Moreover, importantly, the proposed solution takes into account that the sum of the bits transmitted to a given user cannot be higher than the number of bits in its buffer 	in order to avoid excessive allocation of the resources. Therefore, although the users with higher weights can contribute more to increase the objective function (weighted sum-rate), their contribution is limited by their respective buffers' length.} 
\makeatletter 
\if@twocolumn
\begin{figure}[!t]
	\centering
	\includegraphics[width=0.8\columnwidth]{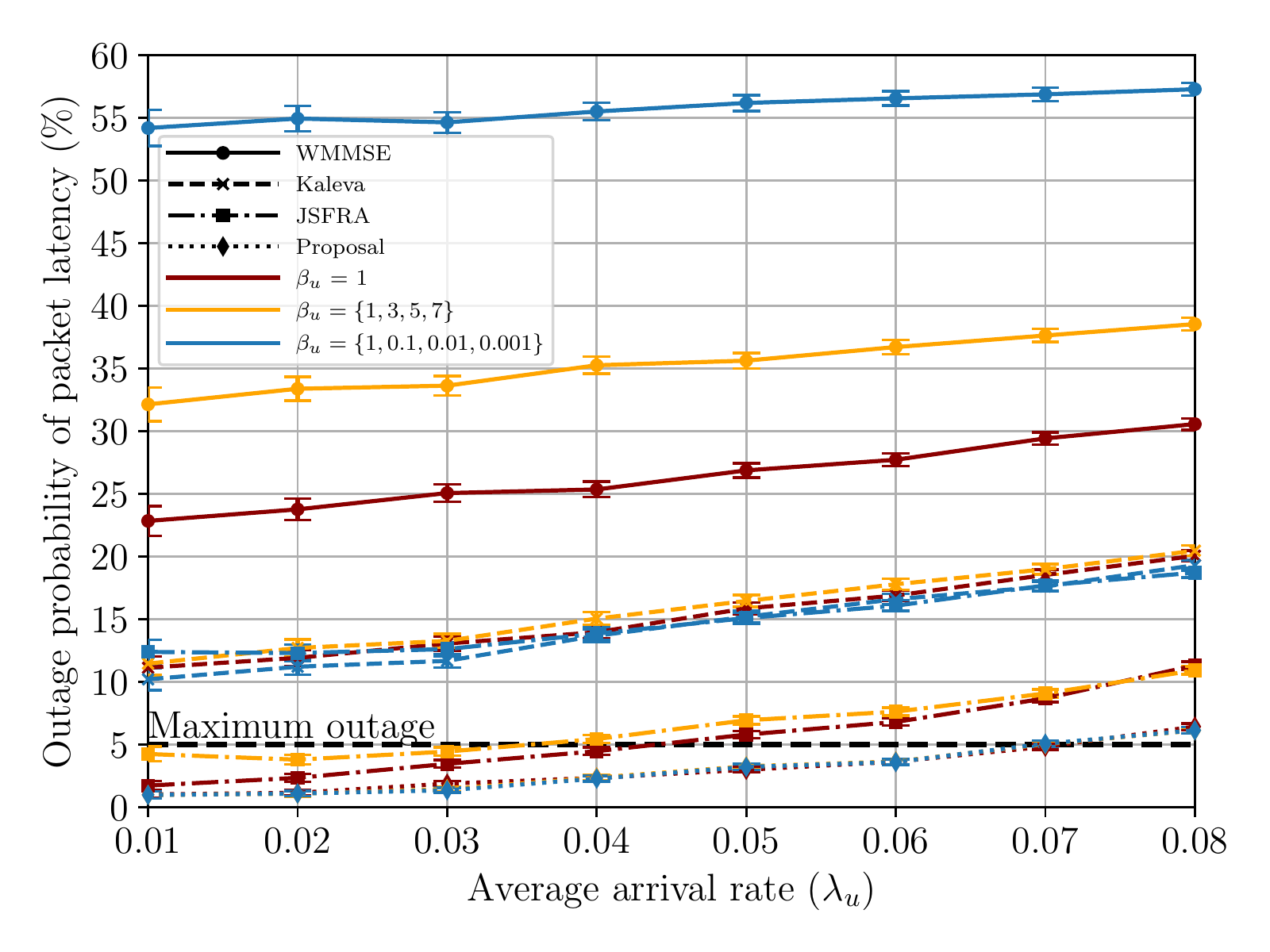}
	\caption{Performance analysis regarding the outage probability with $\{B, U, U_b, N, N_\text{T}, N_\text{R}\} = \{4, 16, 4, 4, 8, 2\}$.}
	\vspace{-2ex}
	\label{FIG:Outage}
\end{figure}
\else
\begin{figure}[!t]
	\centering
	\includegraphics[width=0.5\columnwidth]{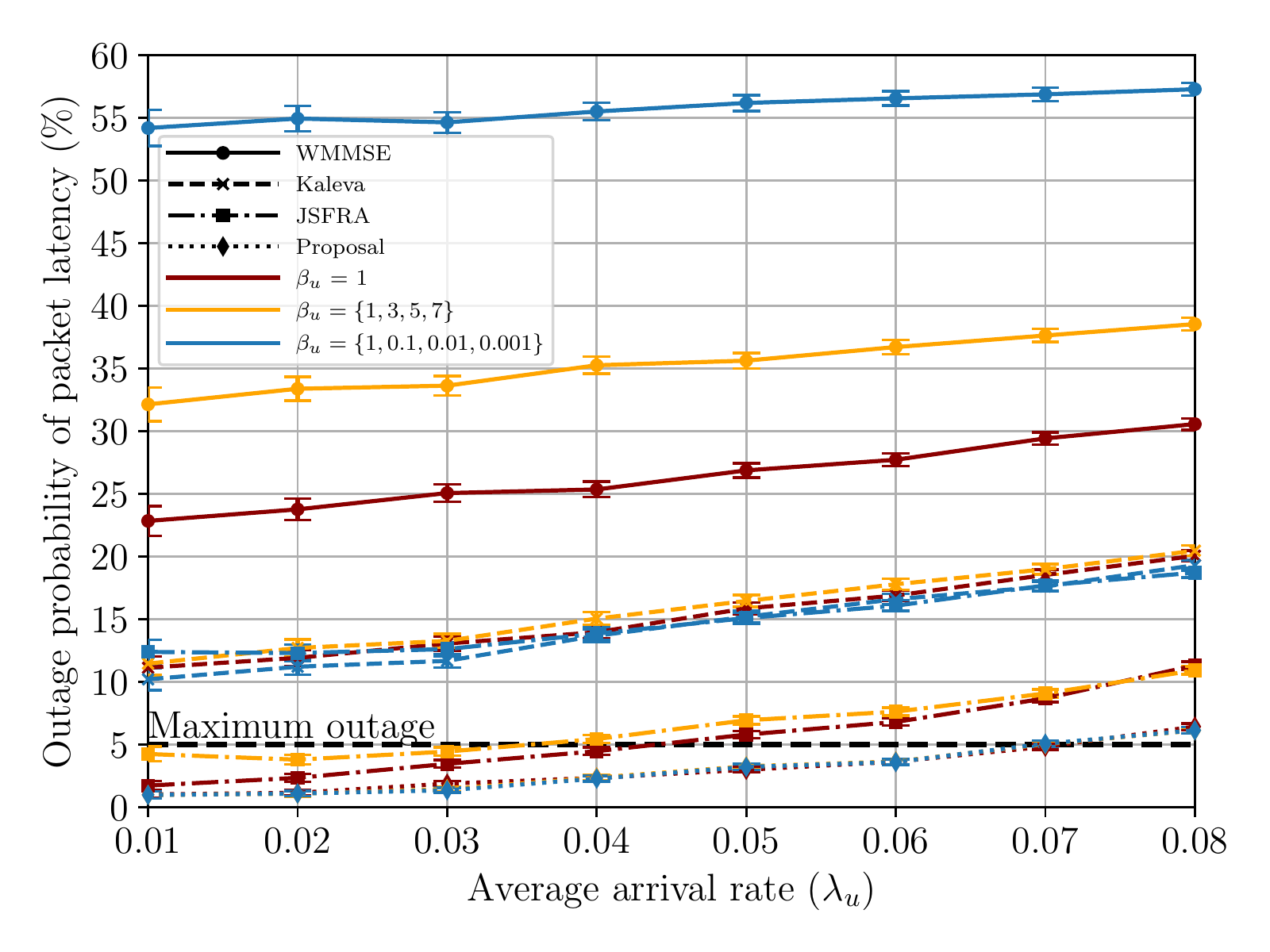}
	\caption{Performance analysis regarding the outage probability with $\{B, U, U_b, N, N_\text{T}, N_\text{R}\} = \{4, 16, 4, 4, 8, 2\}$.}
	\vspace{-2ex}
	\label{FIG:Outage}
\end{figure}
\fi \makeatother

In addition, one can see that the \ac{WMMSE} and Kaleva algorithms present the worst performance in terms of outage among the analyzed solutions. The main reason is that these solutions assume a full-buffer model. Furthermore, the \ac{WMMSE} solution prioritizes the users with best channel conditions without considering minimum per-user rate requirements, thus, increasing the number of outages for packets of users in worst channel conditions. The Kaleva solution, on the other hand, takes into account minimum per-user rate requirements, which makes its performance better than the one of the \ac{WMMSE} solution.
Nevertheless, {the outage probability obtained by the Kaleva solution is still higher than the maximum outage probability allowed.} This occurs because, according to {Proposition~1}, it tries to fulfill the minimum requirements for all users, even though some users do not have bits to be transmitted.

We can also see that the \ac{JSFRA} solution is able to maintain low outage probability values for low values of $\lambda_u$ {and the users' priorities are relatively the same.} The reason is that the \ac{JSFRA} solution aims to minimize the total number of backlogged packets in each \ac{TTI}, which indirectly focuses on delay constraints. {However, in general,} this solution prioritizes users with a higher number of bits in their buffers before considering the users with a smaller number of bits, thus, when $\lambda_u$ increases, users with worst channel conditions tend to accumulate more bits and, consequently, are prioritized. However, those users are not able to achieve high data rates, which increases the waiting time of packets from those users, causing outages.

Finally, the proposed solution is able to maintain the outage probability below the maximum allowed value for almost the entire simulated range of $\lambda_u$.
Unlike the Kaleva solution, the proposed algorithm focuses only on users with buffers that are not empty, thus, it can dedicate more resources to those users and fulfill their minimum rate requirements, which is sufficient to guarantee a low outage.
Note that by guaranteeing the minimum data rate for each user, the proposed algorithm overcomes the problems related to the \ac{JSFRA} solution.
In fact, in the first setup, the \ac{JSFRA} solution maintains the outage probability less than the maximum allowed value for $\lambda_u \leq 0.045$ while the proposed solution is able to fulfill the outage requirement for $\lambda_u \leq 0.072$, yielding a gain of 60\% in terms of supported load.
Moreover, for $\lambda_u = 0.06$, the proposed solution presents a reduction of 43\% of the outage rate compared to the \ac{JSFRA} solution. {Note that these gains compared to the state-of-art algorithms increase for the case in which $\beta_{u} = \{1,\  0.1,\ 0.01,\ 0.001\}$.}

\makeatletter 
\if@twocolumn
\begin{figure}[!t]
	\centering
	\includegraphics[width=0.8\columnwidth]{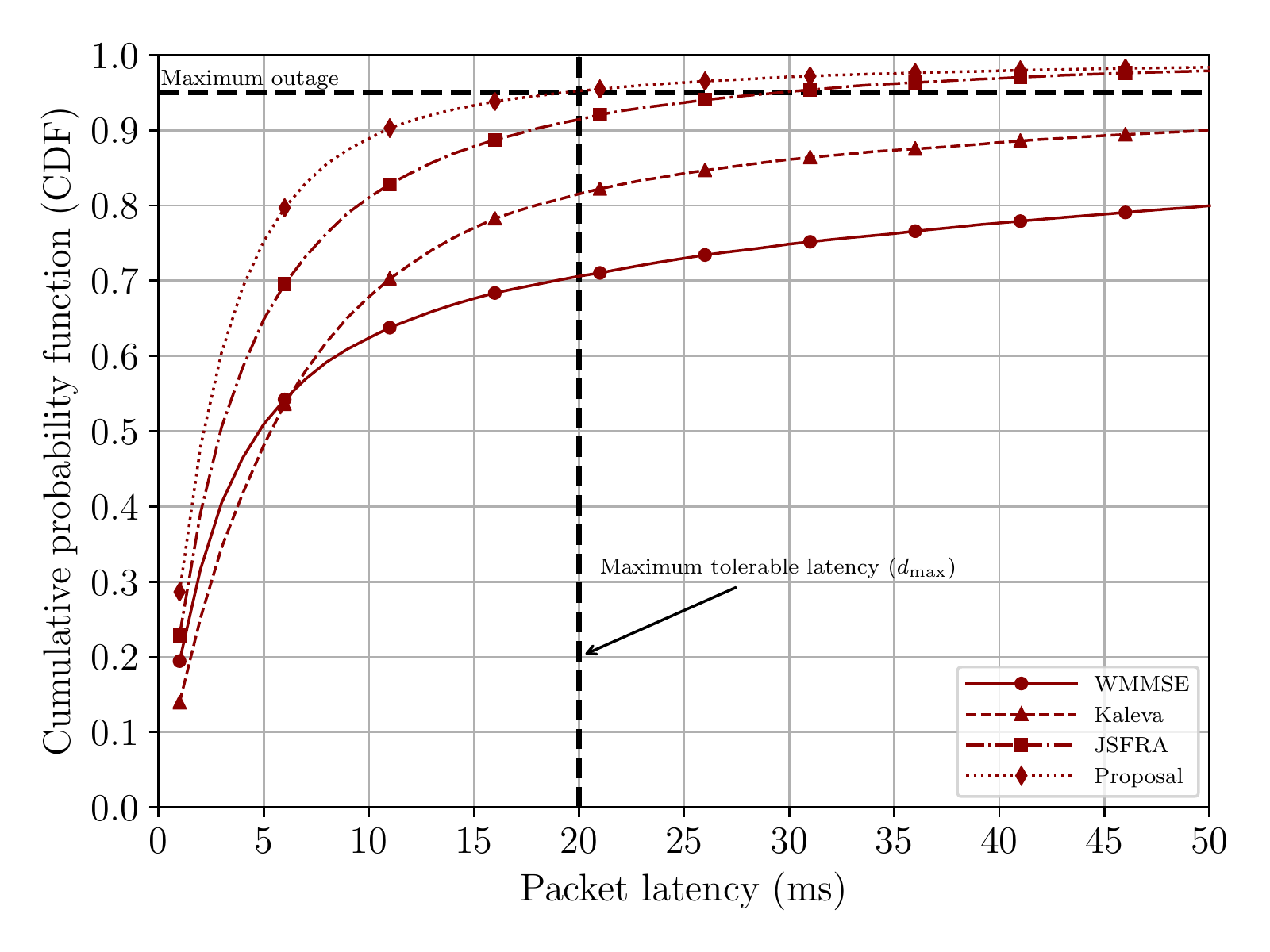}
	\caption{CDF of packet latency for all algorithms using a Poisson traffic model with parameters $\{B, U, U_b, N, N_\text{T}, N_\text{R}, \beta_u,\lambda_u\} = \{4, 16, 4, 4, 8, 2, 1, 0.07\}$.}
	\vspace{-2ex}
	\label{FIG:CDF}
\end{figure}
\else
\begin{figure}[!t]
	\centering
	\includegraphics[width=0.5\columnwidth]{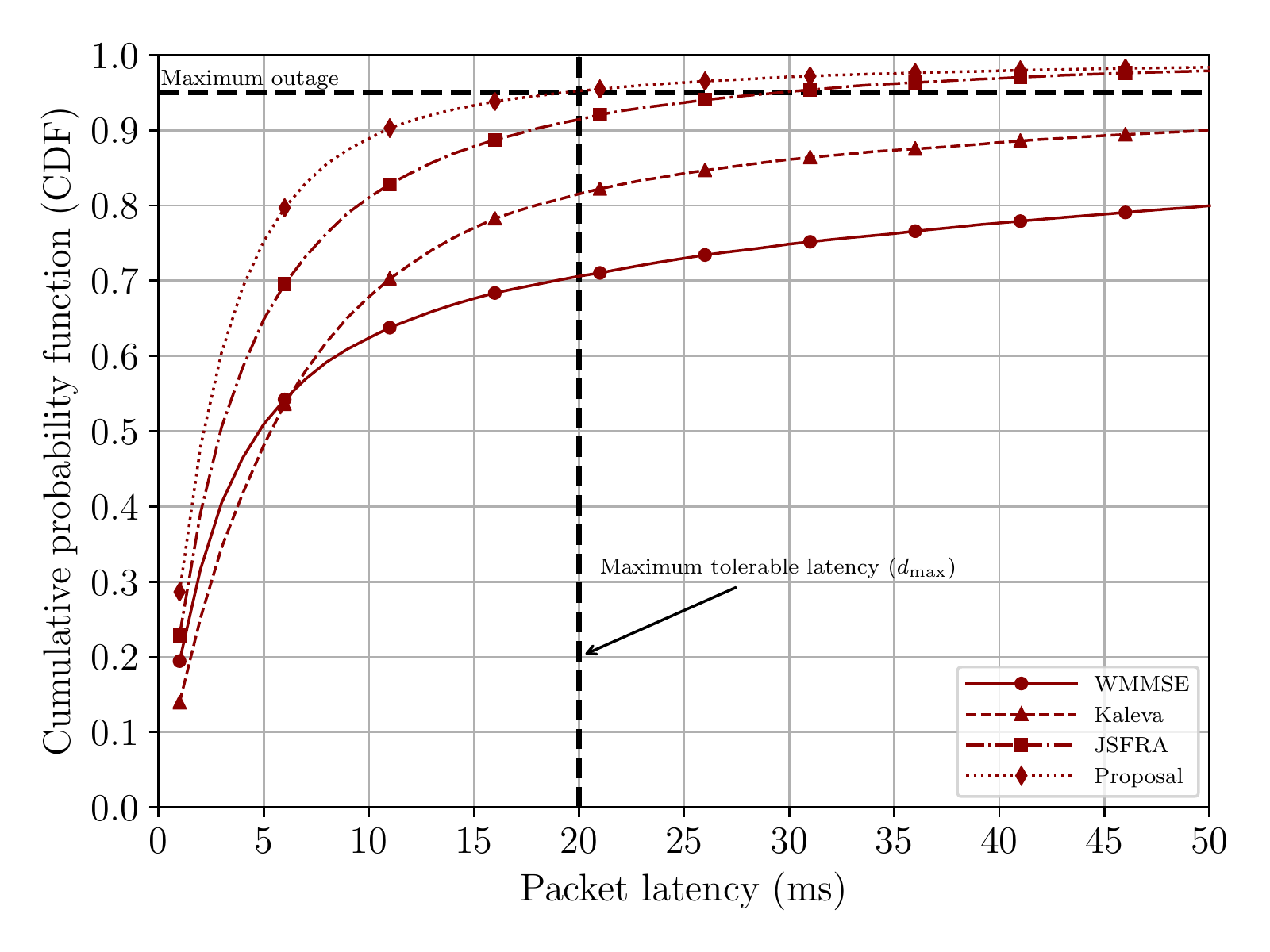}
	\caption{CDF of packet latency using a Poisson traffic model with parameters $\{B, U, U_b, N, N_\text{T}, N_\text{R}, \beta_u,\lambda_u\} = \{4, 16, 4, 4, 8, 2, 1, 0.07\}$.}
	\vspace{-2ex}
	\label{FIG:CDF}
\end{figure}
\fi \makeatother

Up to this point, we have shown that the proposed solution is able to guarantee the outage probability for several packet arrival rates.
Although this indicates that the packet latency is within the allowed range, it does not provide details about the latency of the packets.
Therefore, in \FigRef{FIG:CDF} we present the \ac{CDF} of the latency of the packets for all solutions with $\lambda_u = 0.07$ and $\beta_{u} = 1$ for all users.
As we can see, the curve of the proposed solution is more to the left, which means that it achieves the lowest latencies.
In fact, we can observe that, approximately, 80\% of the packet latencies obtained by the proposed solution are mainly distributed within 1 and 6 ms, while the percentage of packets with latency higher than the maximum tolerable latency is strictly less than the maximum  outage probability allowed, which means that the proposed solution can satisfy the latency requirements.
In addition, at the 50th and 90th percentiles, the proposed solution presents gains of  27\% and 39\% compared to the \ac{JSFRA} solution, respectively.

\subsection{Bursty Traffic Model}
\label{SEC:BURSTY_TRAFFIC_MODEL}

\makeatletter 
\if@twocolumn
\begin{figure}[!t]
	\centering
	\includegraphics[width=0.8\columnwidth]{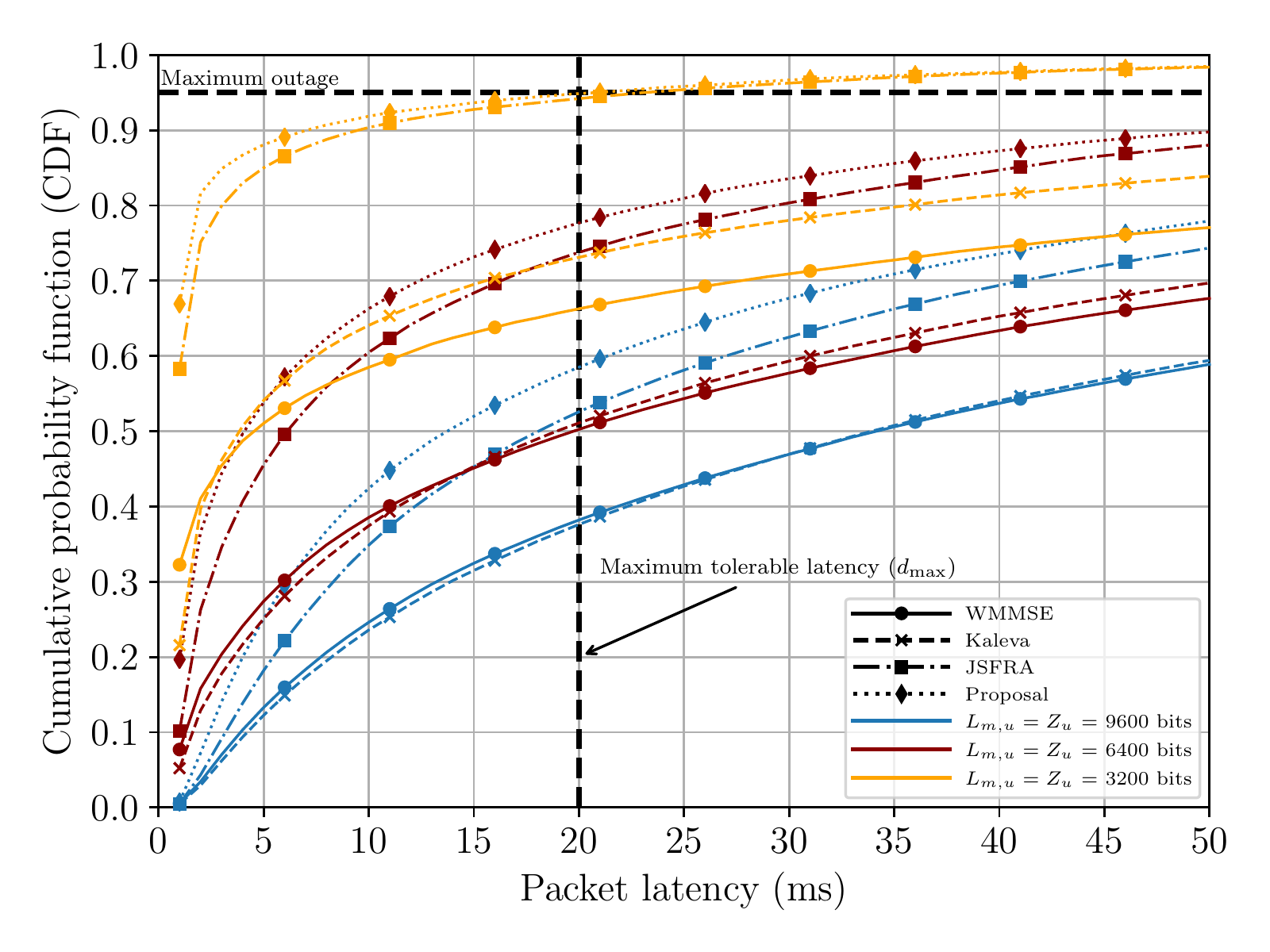}
	\caption{CDF of packet latency for all algorithms assuming bursty traffic scenario with parameters $\{B, U, U_b, N, N_T, N_R, B_u\} = \{4, 16, 4, 4, 8, 2, 1\}$.}
	\vspace{-2ex}
	\label{FIG:CDF_BURSTY}
\end{figure}
\else
\begin{figure}[!b]
	\centering
	\includegraphics[width=0.5\columnwidth]{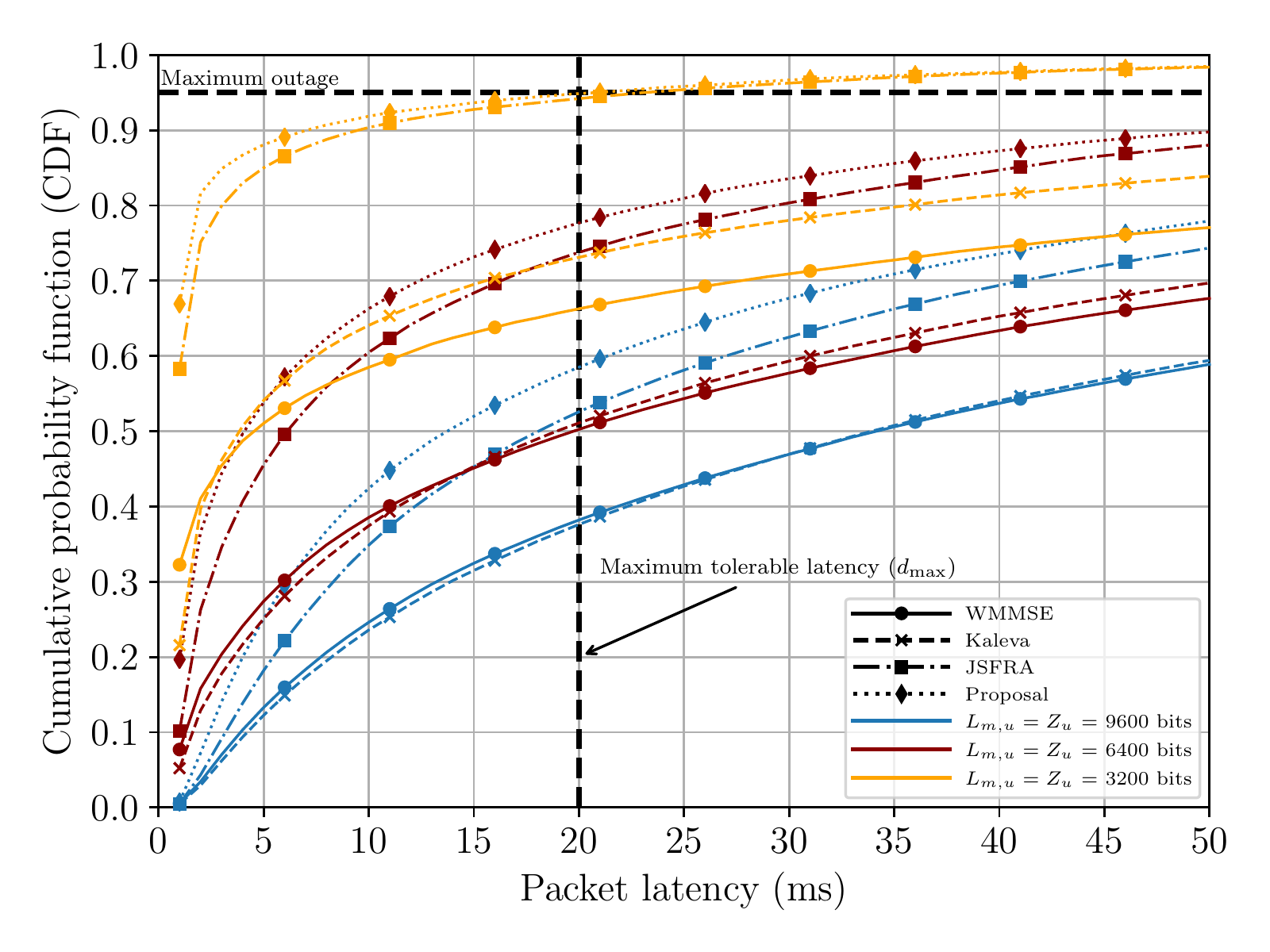}
	\caption{CDF of packet latency assuming bursty traffic scenario with parameters $\{B, U, U_b, N, N_T, N_R, B_u\} = \{4, 16, 4, 4, 8, 2, 1\}$.}
	\vspace{-2ex}
	\label{FIG:CDF_BURSTY}
\end{figure}
\fi \makeatother

In some applications, the arrival of packets is bursty,
which cannot be captured by the Poisson traffic model. 
Thus, in this section, we show how our proposed decentralized solution and the comparison algorithms perform in the presence of a bursty traffic model. For that purpose, we consider a semi-Markov ON-OFF process \cite{Rastegar2020,Liu2017} to model the bursty traffic, where the data traffic pattern is assumed to be i.i.d. among the users.  
Moreover, we use a Pareto random variable to model the duration of the ON state of user $u$, denoted by $\Delta_{u}^{\text{ON}}$, where $\mu_u$ and $\omega_u$ are the shape and scale parameters, respectively. 
The shape parameter $\mu_u$ determines the slope of the Pareto \acl{PDF}, such that increasing values of $\mu_u$ decrease the variance of the variable and concentrate it near 0. Similar to \cite{Rastegar2020}, we set $\mu_u$ equal to~2. The scale parameter $\omega_u$, in its turn, represents the minimum value of the Pareto random variable, which means that the ON state duration would be definitely greater than $\omega_u$ time slots.  In our simulations, we set $\omega_u$ to be equal to 10. Also, we model the duration of the OFF state of user $u$, denoted by $\Delta_{u}^{\text{OFF}}$, using an exponential random variable with $\delta_u$ representing the rate of the exponential distribution, which is set to 100 during the simulation. 
In addition, during the ON state, user $u$ continuously receives one packet per time slot with size equal to $Z_u$ bits, while during the OFF state it is idle and has no data to receive.
Using this
model, to compute the minimum data rate requirement using Proposition 1, we set $\lambda_u$ and $\bar{L}_{u,m}$ equal to 1 and $Z_u$, respectively.

In Fig.~\ref{FIG:CDF_BURSTY}, we present a CDF of the packet latency for all algorithms assuming the bursty traffic scenario. The packet latency increases as the packet size increases for all solutions, which is an expected behavior since with higher packet sizes the traffic becomes more intense. The proposed solution presents the lowest values of packet latency compared to the other solutions considering all traffic loads.  Note that the proposed solution fails to meet the outage requirements when the packet sizes are equal to 6400 and 9600 bits. The reasoning for this is that, during the ON state, the arrival rate of packets is constant, thus, the minimum data rate requirement is increased according to Proposition 1. Then, fulfilling these rate demands becomes even more difficult, which leads to a large number of backlogged packets and, consequently, higher values of outage. Even in this situation, we observe that for small packet sizes, the proposed solution is able to fulfill the outage demands, which shows that the proposed solution could be applied for low-intensity bursty traffic scenarios.

\subsection{Imperfect Channel State Information}
\label{SEC:IMPERFECT_CSI}

During the derivation of Algorithms 1 and 2, we assumed that perfect \ac{CSI} is available at the transmitters and receivers, which can be very difficult to be obtained
in practical systems. In this subsection, we analyze the performance of the proposed distributed solution (i.e., Algorithm 2) and comparison algorithms under imperfect \ac{CSI}. For that, we modeled the \ac{CSI} imperfection by assuming that the \acp{BS} estimate the channel using an \ac{MMSE} estimator. Thus, the estimated channel matrix satisfies \cite{rusek2013scaling, braga2020}: {$
	\hat{\mtH}_{b_{i},u,n} = \varrho\mtH_{b_{i},u,n} + \sqrt{1 - \varrho^2}\bm{\Gamma}$, 
where $\bm{\Gamma}\in\bbC^{N_R\times N_T}$} is an error matrix with complex Gaussian i.i.d. entries with zero mean and unit variance, while ${0 \leq \varrho \leq 1}$ denotes the reliability of the channel estimation. The $\varrho$ parameter is set in such a way that the \ac{MSE} between the estimated channel matrix and the actual one is approximately -10~dB and -5.7~dB, reflecting different reliability scenarios.

Fig.~\ref{FIG:IMPERFECT_CSI} presents the \ac{CDF} of the values of packet latency for all analyzed solutions considering different levels of \ac{CSI} imperfection. Again, the curve of the proposed solution is more to the left, thereby, it achieves the lowest values of packet latency.  In addition, we have that the JSFRA, Kaleva and the WMMSE solutions are drastically affected by channel estimation errors.  Meanwhile, we can see that the proposed solution is still able to fulfill the demands of the system when the channel estimation errors are low. Indeed, when the \ac{MSE} is equal to -10 dB, the percentage of packets with latency higher than the maximum tolerable latency is strictly less than the maximum allowed outage probability.
Thus, even under the effects of channel estimation errors, the proposed solution meets the data rate requirements and, consequently, the outage probability. This is an interesting result because the proposed solution does not consider the channel errors in its modeling and even so it can satisfy the outage requirements. When the channel estimation error increases, i.e., MSE = -5.7 dB, we observe that the outage probability of the proposed solution increases to, approximately, 11\%, which is higher than the maximum allowed. However, the proposed solution achieves a gain of 38\% compared to the JSFRA solution in this situation. Improving the performance of the proposed solution with imperfect \ac{CSI} is out of the scope of this paper and is left as a perspective for future works.

\makeatletter 
\if@twocolumn
\begin{figure}[!t]
	\centering
	\includegraphics[width=0.8\columnwidth]{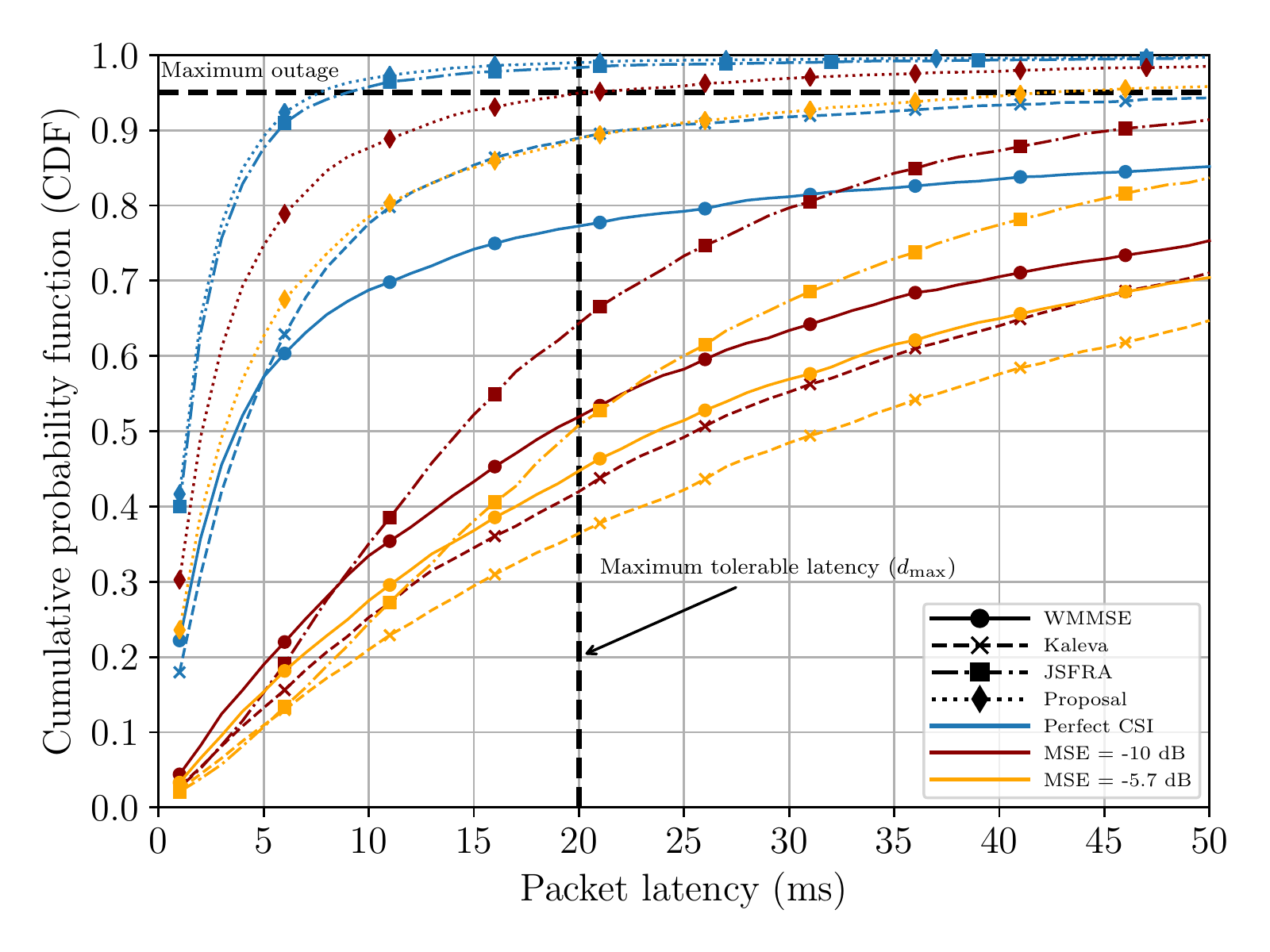}%
	\caption{
		CDF of packet latency for all algorithms with different levels of CSI and parameters $\{B, U, U_b, N, N_T, N_R, \beta_u, \lambda_u\} = \{4, 16, 4, 4, 8, 2, 1, 0.01\}$.}
	\label{FIG:IMPERFECT_CSI}
\end{figure}
\else
\begin{figure}[!h]
	\centering
	\includegraphics[width=0.5\columnwidth]{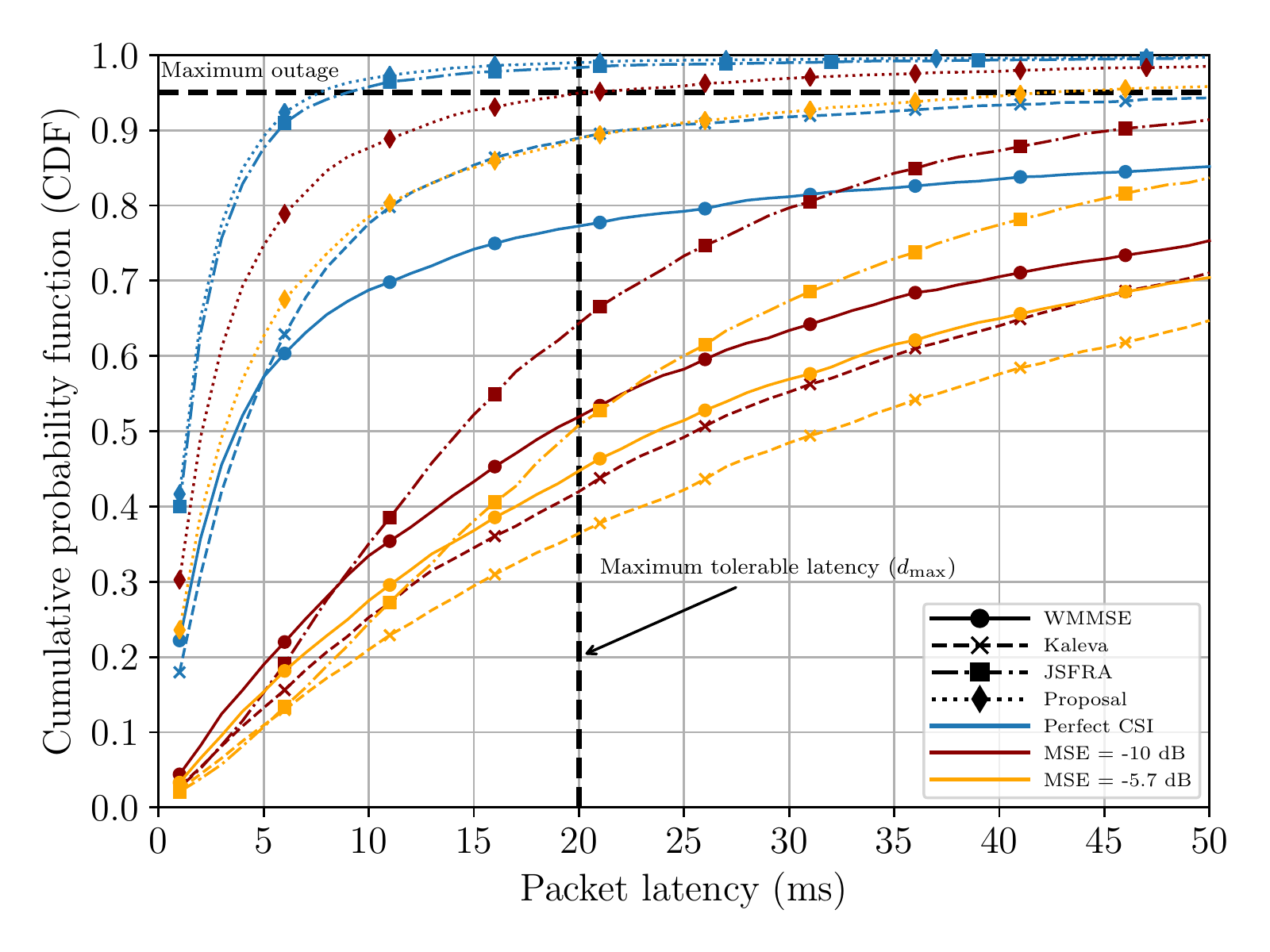}%
	\caption{
		CDF of packet latency with different levels of CSI imperfection and parameters $\{B, U, U_b, N, N_T, N_R, \beta_u, \lambda_u\} = \{4, 16, 4, 4, 8, 2, 1, 0.01\}$.}
	\label{FIG:IMPERFECT_CSI}
\end{figure}
\fi \makeatother

\section{Conclusions}
\label{SEC:CONCLUSIONS}

In this work, we investigated a variant of the weighted sum-rate maximization problem subject to latency outage probability constraints in multicell \ac{MU-MIMO} \ac{OFDM} systems with a finite buffer model. The initially formulated problem was verified to be non-convex and very difficult to be optimally solved.  It was then reformulated and solved, iteratively, up to a locally optimal point by using the max-plus queuing method from network analysis, the well known \ac{MSE}-\ac{SINR} relation when using \ac{MMSE}, as well as \ac{SCA}. In addition, a decentralized solution with relaxed feasible initialization requirements was proposed based on the dual decomposition and Lagrangian relaxation of the rate constraints. Also, signaling aspects for practical implementation of the decentralized solution and a detailed convergence analysis were provided.
	
Unlike previous works, a more realistic channel model was utilized with space, frequency and time correlations. The numerical results showed that the proposed framework outperforms state-of-the-art algorithms in terms of outage probability and latency for different scenarios. Indeed, compared to the benchmarking solutions, the proposed solution presented a reduction of approximately 43\% of outage probability and a gain of 60\% in terms of the supported load in scenarios where users have equal user weights. {Thus, we conclude that the proposed solutions present the currently available best performance
to the stated problem considering the existing methods to solve such problems.} Finally, as perspective for further studies, we indicate the development of solutions that take into account channel estimation and extensions of the proposed framework using other traffic models.
	
\section*{Appendix A \\ Convergence Analysis}

In this appendix, we perform the converge analysis for both centralized and decentralized algorithms, which is based on~\cite{Kaleva2016}. We assume a sufficient number of inner subgradient iterations for both centralized and decentralized algorithms. In other words, in order to guarantee monotonic improvement with respect to the objective function after each transmit beamformers iteration, Algorithms \ref{ALG:SCA_ALGORITHM} and \ref{ALG:DEC_ALG_SCA} perform enough subgradient updates~\cite{Boyd2004, Bertsekas1999}.

\begin{remark}
	Several iterations of the dual decomposition and \ac{SCA} updates can be performed in the inner loops of Algorithms \ref{ALG:SCA_ALGORITHM} and \ref{ALG:DEC_ALG_SCA} during the updates of the variables {$\{\hat{r}_{u,n}\}_{\forall (u\in\stU_t,n)}$} and the transmit beamformers {$\{\mtM_{u,n}\}_{\forall (u\in\stU_t,n)}$}. The number of inner updates should guarantee the monotonicity regarding the global objective function.
	\label{REMARK:MONOTONICITY}
\end{remark}

We follow our analysis by showing that the feasible set of \eqref{EQ:PROB_MSE} is compact.
Since the power constraints in \eqref{PROB:MAX_RATE_LATENCY_CONS1} are convex and compact, we have that the feasible set of the transmit beamformers {$\{\mtM_{u,n}\}_{\forall (u\in\stU_t,n)}$} is convex and compact.
Analogously, the feasible set of the variables {$\{\hat{r}_{u,n}\}_{\forall (u\in\stU_t,n)}$} is also convex and compact.
Moreover, since the noise power is non-zero, the receive covariance matrix in \eqref{EQ:MMSE} is always invertible and, consequently, the mapping between the transmit and receive beamformer is a continuous map \cite{Boyd2004}.
Based on that the set comprising the feasible \ac{MMSE} receive beamformers {$\{\mtW_{u,n}\}_{\forall (u\in\stU_t,n)}$} is closed and bounded, which shows that such a set is also compact.
Finally, we can represent the updates iterations of the variables {$\{\hat{r}_{u,s,n}\}_{\forall (u\in\stU_t,s,n)}$}, as well as the receive and transmit beamformers, using infimal maps \cite{Kaleva2016,Dantzig1966}.
According to~\cite{Dantzig1966}, since the set of all optimization variables is compact, as stated before, the infimal maps modeling the updates of the optimization variables are closed point-to-set maps.

\begin{prop}
	The objective function of problem \eqref{EQ:PROB_MSE} is monotonic and converges with Algorithm \ref{ALG:SCA_ALGORITHM}. \label{PROP:MONOTONIC}
\end{prop}
\begin{IEEEproof}
	The \ac{MMSE} receivers are the unique rate optimal receivers for problem~\eqref{EQ:PROB_MSE} in the sense that they maximize the per-stream \ac{SINR}, i.e., they maximize the rate for each user \cite{Kaleva2016, Pennanen2016}. Then, we can conclude that the objective of problem \eqref{EQ:PROB_MSE} is strictly increasing for each receive beamformer update given
	by \eqref{EQ:MMSE}. Also, it was shown in \cite{Marks1978} that the solution of the \ac{SCA} subproblem in Algorithm~\ref{ALG:SCA_ALGORITHM} is either a solution of the original problem or the objective is monotonically improved. Furthermore, given Remark~\ref{REMARK:MONOTONICITY}, the monotonicity is extended to Algorithm~\ref{ALG:DEC_ALG_SCA}.
	Finally, since the objective is bounded by the power and rate constraints, we can claim the convergence of the objective function in problem~\eqref{EQ:PROB_MSE} when executing Algorithm~\ref{ALG:SCA_ALGORITHM} and Algorithm~\ref{ALG:DEC_ALG_SCA}.
\end{IEEEproof}

Once we have shown that the algorithms can be modeled as closed infimal maps and are monotonic with respect to the objective function of problem~\eqref{EQ:PROB_MSE}, it follows from the  convergence theorem in~\cite{Zangwill1969} that the sequence of iterates generated by Algorithm~\ref{ALG:SCA_ALGORITHM} and Algorithm~\ref{ALG:DEC_ALG_SCA} has at least one accumulation point and each accumulation point is a generalized fixed point.
	

However, we can make the convergence results stronger and show that the Algorithms 1 and 2 converge to a unique solution for all fixed points. Indeed,  we can obtain this behavior by using a uniquely defined generalized inverse operation, such as the Moore-Penrose pseudoinverse, in \eqref{EQ:TRANSMIT_BEAMFORMERS} \cite{Kaleva2016}.
Nevertheless, we have that the set of the feasible fixed points is infinite, since there is an \ac{SINR} equivalence for different complex beamformers with some different phase rotation, consequently, convergence to a single fixed point cannot be guaranteed \cite{Meyer1976}.

Then, we should show that any fixed point of Algorithm \ref{ALG:SCA_ALGORITHM} is a \ac{KKT} point of problem \eqref{PROB:MAX_RATE_LATENCY_REF}.
\begin{prop}
	Any fixed limit point {$\{\mtW_{u,n}^{\ast}, \mtM_{u,n}^{\ast}, \hat{r}^{\ast}_{u,n}, \mtE_{u,n}^{\ast}\}_{\forall (u\in\stU_t,n)}$} of Algorithm \ref{ALG:SCA_ALGORITHM} is a \ac{KKT} point of problem \eqref{PROB:MAX_RATE_LATENCY_REF}.
\end{prop}
\begin{IEEEproof}
	Based on \cite{Marks1978}, where it was shown that the \ac{SCA} algorithm stops at a \ac{KKT} point, or the limit of any convergent sequence is a \ac{KKT} point, with a slight difference due to the extra step involving the receive beamformer updates. That said, we have that the primal and dual constraints always hold for problem \eqref{PROB:MAX_RATE_LATENCY_REF}, since the convex approximation is only applied for the constraints \eqref{EQ:TAYLOR}.
	Consequently, we can focus only on the constraints affected by \ac{SCA}.
	
	Let us start by defining {$\varUpsilon(\mtE_{u,n},\mtE_{u,n}^{(k)})$} as the first-order Taylor approximation around a fixed \ac{MSE} point in \eqref{EQ:TAYLOR}. Thus, from the convergence to a fixed point and definition of the first-order linear approximation, we have that {$-\log_2 \det\left(\mtE_{u,n}^{\ast}\right) = \varUpsilon(\mtE_{u,n}^{\ast},\mtE_{u,n}^{\ast})$} and {$\frac{-\partial \log_2\det\left(\mtE_{u,n}^{\ast}\right)}{\partial \mtE_{u,n}} = \frac{\partial\ \varUpsilon(\mtE_{u,n}^\ast,\mtE_{u,n}^\ast) }{\partial \mtE_{u,n}},\ \forall(u\in\stU_t,n)$}.
	Thus, by definition, we have that the first-order optimality conditions hold. In addition, the \ac{MSE} as well as transmit and receive beamformers in \eqref{EQ:MSE}, \eqref{EQ:TRANSMIT_BEAMFORMERS} and \eqref{EQ:MMSE}, respectively, are directly solved from the first-order optimality conditions, which means that they also satisfy the optimality conditions. Regarding the complementary slackness conditions, it is easy to see that the constraints \eqref{EQ:TAYLOR} hold tight at any
	fixed point, consequently, {$\hat{r}_{u,n} = \varUpsilon(\mtE_{u,n}^\ast,\mtE_{u,n}^\ast) \Rightarrow \psi_{u,n}^{\ast}(\hat{r}_{u,n}^{\ast} - \varUpsilon(\mtE_{u,n}^\ast,\mtE_{u,n}^\ast)) = 0,\ \forall(u\in\stU_t,n)$}. Similar analyses also apply for the rate and \ac{MSE} constraints. In addition, we can also observe that, since the linear
	approximation generates a lower bound for convex functions, the primal feasibility holds, i.e., {$\hat{r}_{u,n}^{\ast} \leq \varUpsilon(\mtE_{u,n}^\ast,\mtE_{u,n}^\ast) \leq -\log_2\det\left(\mtE_{u,n}^{\ast}\right)$}. Then, we can conclude that any fixed point is also a \ac{KKT} point of problem \eqref{PROB:MAX_RATE_LATENCY_REF}, which was also
	observed in \cite{Marks1978}. The equivalence between a \ac{KKT} point of \eqref{PROB:MAX_RATE_LATENCY_REF} and \eqref{EQ:PROB_MSE} follows directly from the well-known relation between \ac{MSE} and \ac{SINR} when using \ac{MMSE} receivers \cite{Christensen2008,Shi2011}, for which a similar rigorous proof is shown in \cite{Shi2011}.		
\end{IEEEproof}

\bibliographystyle{IEEEtran}
\bibliography{IEEEabrv,biblio}

\end{document}